\documentclass{article}
\usepackage{fullpage}
\usepackage{graphicx}           % Standard graphics package
\usepackage{subfigure}          % subfigure package
\usepackage{amsmath}            % For \numberwithin
\usepackage{amssymb}            % For, e.g., \mathbb, \big, and \mapsto
\usepackage{url}                % For email addresses, hypertext links, ...
\usepackage{ifpdf}              % For pdf and postscript files
\usepackage{pdfpages}           % For inserting pdf pages
\usepackage{algorithm}
\usepackage[noend]{algpseudocode}
\usepackage{multirow}
\usepackage{setspace}

\pdfoutput=1

\ifpdf
\DeclareGraphicsExtensions{.pdf,.png,.jpg}
\else
\DeclareGraphicsExtensions{.eps,.ps}
\fi

\begin{document}

%\begin{frontmatter}

\title{Graph Coloring Algorithms for Multi-core and Massively Multithreaded Architectures}

\author{
\"Umit V. \c{C}ataly\"urek\footnote{Departments of Biomedical Informatics and Electrical \&
  Computer Engineering, The Ohio State University. {\tt umit@bmi.osu.edu}}
\and 
John Feo\footnote{Pacific Northwest National Laboratory. {\tt John.Feo@pnl.gov}}
\and 
Assefaw H. Gebremedhin\footnote{Department of Computer Science, Purdue University. {\tt agebreme@purdue.edu}}
\and 
Mahantesh Halappanavar\footnote{Pacific Northwest National Laboratory. {\tt Mahantesh.Halappanavar@pnl.gov}}
\and 
Alex Pothen\footnote{Department of Computer Science, Purdue University. {\tt apothen@purdue.edu}}
}

\maketitle

\begin{abstract}
We explore the interplay between {\em architectures} and {\em algorithm design}
in the context of shared-memory platforms and a specific graph problem of
central importance in scientific and high-performance computing,
distance-1 graph coloring.
We introduce two different kinds of  multithreaded heuristic algorithms
for the stated, NP-hard, problem.
The first algorithm relies on {\em speculation} and {\em iteration}, 
and is suitable for {\em any} shared-memory system.
The second algorithm uses {\em dataflow} principles, and is targeted at
the non-conventional, massively multithreaded Cray XMT system.  
We study the performance of the algorithms on the Cray XMT and
two multi-core systems, Sun Niagara 2 and Intel Nehalem.
Together, the three systems represent a spectrum of multithreading capabilities
and memory structure. 
As testbed, we use synthetically generated large-scale graphs carefully 
chosen to cover a wide range of input types.
The results show that the algorithms have scalable runtime performance
and use nearly the same number of colors as the underlying 
serial algorithm, which in turn is effective in practice.
The study provides insight into the design of high performance algorithms
for irregular problems on many-core architectures.

\end{abstract}

\section{Introduction}
\label{sec:intro}

Graph problems  frequently arise in many practical applications, including
computational science and engineering, data mining, and data analysis.
When the applications are large-scale, solutions 
need to be obtained on parallel computing platforms.
High performance and good scalability are hard-to-achieve on
graph algorithms, for a number of well-recognized reasons 
\cite{DBLP:journals/ppl/LumsdaineGHB07}:
Runtime is dominated by memory latency rather than processor speed, and
typically there is little (or no) computation involved to hide memory access costs. 
Access patterns are irregular and are determined by the structure of the input graph,
rendering prefetching techniques inapplicable. 
Data locality is poor, making it difficult to obtain good memory system performance. 
While concurrency can be abundant, it is often fine-grained, requiring concurrent processes to
synchronize at individual vertices or edges.

For these reasons, graph algorithms that perform and scale well on
distributed memory machines are relatively small in number and kind. 
More success stories have been reported on shared memory platforms, 
and interest in these platforms is growing with the increasing preponderance, 
popularity, and sophistication of {\em multi-core} architectures 
\cite{Kurzak+,Bader+:Multicore-graphalgs,Madduri+:Multithreaded-graphalgs}. 

The primary mechanism for tolerating memory latencies on most
shared memory systems is the use of {\em caches}, but caches have been found rather 
ineffective for many graph algorithms. A more effective mechanism is {\em multithreading}. 
By maintaining multiple threads per core and switching among them in the event of a long
latency operation, a multithreaded processor uses parallelism to hide latencies. 
Whereas caches ``hide'' only memory latencies, thread parallelism can hide both 
memory and synchronization overheads. 
Thus, multithreaded, shared memory systems are more suitable
platforms for many graph algorithms than either distributed memory machines or 
single-threaded, multi-core shared memory systems. 

We explore in this paper the interplay between architectures and algorithm design
in the context of shared-memory multithreaded systems and a specific graph problem, 
{\em distance-1 graph coloring}.

Graph coloring in a generic sense is an abstraction for partitioning
a set of binary-related objects into subsets of independent objects. A need for such a
partitioning arises in many situations where there is a scarce resource that needs to be
utilized optimally. 
One example of a broad area of application is  
in discovering concurrency in parallel computing, 
where coloring is used to identify subtasks that can be carried out or data elements
that can be updated simultaneously \cite{jones94scalable-v2,saad96ilum-v2,hysom01}.
On emerging heterogenous architectures,  coloring algorithms on certain ``interface-graphs'' are used at runtime to decompose computation into concurrent tasks that can be mapped to different processing units \cite{Listz}.    
Another example of a broad application area of coloring is the efficient computation of 
sparse derivative matrices \cite{CoMo:83,CoMo:84,GMP05}.

Distance-1 coloring is known to be 
NP-hard not only to solve optimally but also in an approximate sense \cite{Zuckerman}. 
However, {\em greedy} algorithms, which run in linear time
in the size of the graph, often yield near optimal solutions on graphs that arise in practice, 
especially when the greedy algorithm is initialized with careful {\em vertex ordering} techniques
\cite{CoMo:83, colpack-acm}. 
In contexts where coloring is used as a step to enable some overarching computation, 
rather than being an end in itself,  greedy coloring algorithms are attractive alternatives 
to slower, local-improvement type heuristics because they yield sufficiently 
small number of colors while using run times that are much shorter than the computations 
they enable. 

{\bf Contributions. }
This paper is concerned with   the effective parallelization of greedy 
distance-1 coloring algorithms on multithreaded architectures.
To that end, we introduce two different kinds of multithreaded algorithms  targeted at 
two different classes of architectures.

The first multithreaded algorithm is suitable for {\em any} shared memory system, 
including the emerging and rapidly evolving multi-core platforms. 
The algorithm relies on {\em speculation} and {\em iteration}, and
is derived from the parallelization framework for graph coloring on 
distributed memory architectures developed in 
Bozda\u{g} et al.~\cite{BGMBC08,BCGMBO:sisc}. 
We study the performance of the algorithm on three different platforms,
an Intel Nehalem, a Sun Niagara 2, and a Cray XMT. 
These three systems employ multithreading---in varying degrees---to hide latencies; 
the former two additionally rely on cache hierarchies to hide latency. 
The amount of concurrency explored in the study ranges from small 
(16 threads on the Nehalem) to medium (128 threads on the Niagara)
to massive (16,384 threads on the XMT).
We find that the limited parallelism and coarse synchronization of the iterative algorithm fit well
with the limited mulithreading capabilities of the Intel and Sun systems. 

The iterative algorithm runs equally well on the XMT, but it does not take advantage of
the system's massively multithreaded processors and hardware support for fast synchronization.
To better exploit the XMT's unique hardware features, we developed  
a fine-grained, {\em dataflow} algorithm making use of the single-word synchronization
mechanisms available on the XMT.
The dataflow algorithm is the second algorithm presented in this paper. 

The performance study  is conducted using  a set of carefully chosen synthetic graphs 
representing a wide spectrum of input types.
The results show that the algorithms perform and scale well on massive graphs 
containing as many as a billion edges. This is true of the dataflow algorithm on 
the Cray XMT and of the iterative algorithm on all three of the platforms considered;
on the XMT, the dataflow algorithm runs faster and scales better than the iterative algorithm. 
The number of colors the algorithms use is nearly the same as that used
by the underlying serial algorithm.

In the remainder of this section, we briefly discuss related work. 
The rest of the paper is organized as follows.
To establish background, especially for the design of the dataflow algorithm,
we review in Section~\ref{sec:platforms} basic architectural features of the 
three platforms used in the study.
We describe the coloring algorithms in detail and discuss their
relationship with previous work in Section~\ref{sec:algorithms}. 
In Section~\ref{sec:graphs}, we discuss the rationale for, the generation of, 
and the characteristics of the synthetic graphs used in the study.  
The experimental results are presented and analyzed in detail in Section~\ref{sec:results}.
We end by drawing broader conclusions in Section~\ref{sec:conc}.

{\bf Related work. }
Graph coloring has a vast literature, and various approaches have been taken 
to solve coloring problems on computers.
Exact algorithms include those based on integer programming, semidefinite programming, 
and constraint programming.
Heuristic approaches include the greedy algorithms mentioned earlier, local search algorithms,  
population-based methods, and methods based on successively finding independent sets 
in the graph \cite{COLOR03}. 
There have been two DIMACS implementation challenges on coloring \cite{DIMACS}, 
where the focus was on obtaining the smallest number of colors possible for any problem
instance; algorithms merely focused on this objective typically have large running times, 
and as a result they can handle only small instances of graphs, 
often with at most a thousand vertices.

A different approach has been taken by Turner~\cite{Turner}, 
who analyzes a heuristic algorithm due to Br{\'{e}}laz \cite{Brelaz}. 
Turner considers random graphs that have the property that they can be colored with $k$ colors, 
where $k$ is a constant, and shows that Br{\'{e}}laz's algorithm colors such graphs  
with high probability using $k$ colors. 
Further analysis of this phenomenon is provided by Coja-Oghlan, Krivelevich, and 
Vilenchik \cite{Oghlan}. 
An experimental study we conducted in a related work \cite{colpack-acm}
has shown that on many graphs from various application areas,  
a greedy coloring algorithm, initialized with appropriate vertex ordering techniques,  
yields fewer colors than Br{\'{e}}laz's algorithm, while running faster.
A survey of several graph coloring problems (e.g., distance-$1$, distance-$2$, star, and 
acyclic coloring on general graphs; and partial distance-$2$ coloring  on bipartite graphs) 
as they arise in the context of automatic differentiation is provided in 
Gebremedhin et al. \cite{GMP05}. 
Effective heuristic algorithms for the star and acyclic coloring problems and case studies
demonstrating the use of those coloring algorithms in sparse Hessian computation have been 
presented in subsequent works  \cite{GTMP07, GPTW09}.  

\section{Architectures}
\label{sec:platforms}

The three platforms considered in this study---an Intel Nehalem, a Sun Niagara~2, 
and a Cray XMT---represent a broad spectrum of multithreading capabilities and memory structure.
The Intel Nehalem relies primarily on a cache-based hierarchical
memory system as a means for hiding latencies, supporting just
two threads per core.  
The Cray XMT has a flat, cache-less memory system and uses
massive multithreading as the sole mechanism for tolerating latencies
in data access.
The Sun Niagara~2 offers a middle path by using a moderate number
of hardware-threads along with a hierarchical memory system. 
We review in the remainder of this section some of the specific architectural features 
and programming abstractions available in the three systems, focusing on aspects 
relevant to our work on algorithm design and performance evaluation.

\subsection{The Intel Nehalem}
\label{sec:nehalem}

\begin{figure}
\centering
\includegraphics[scale=0.35]{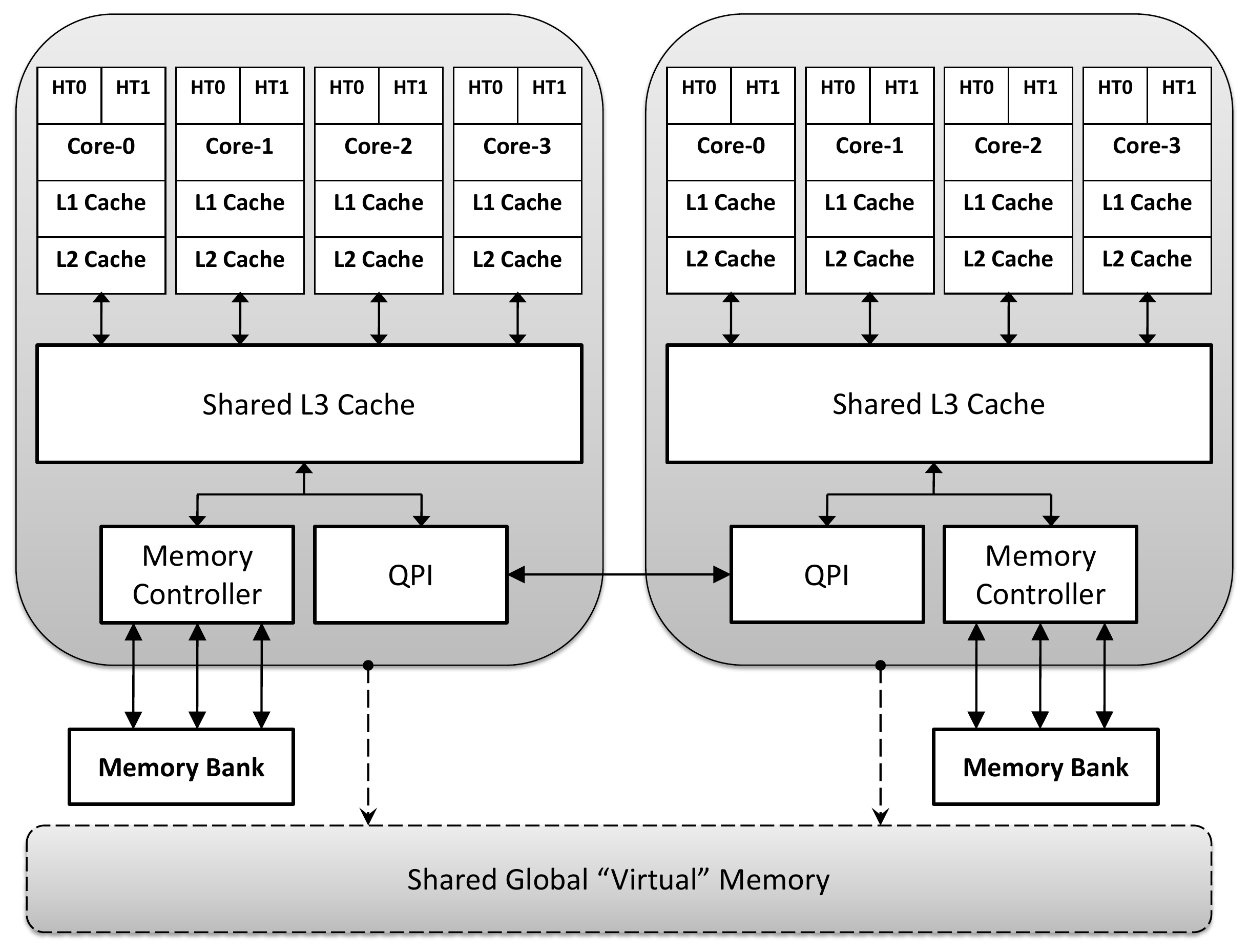}
\caption{Block diagram of the Intel Nehalem.}
\label{fig:nehalem}
\end{figure}

The Intel system we considered, Nehalem EP (Xeon E5540), consists of two quad-core Gainestown processors 
running at $2.53$ GHz base core frequency. See Figure~\ref{fig:nehalem} for a block diagram.
Each core supports two threads and has one instruction pipeline.
The multithreading variant on this platform is {\em simultaneous}, 
which means multiple instructions from ready threads are executed in a given cycle. 

The system has $24$ GB total memory. 
The memory system per core  consists of:
$32$ KB, $2$-way associative I1 (Instruction) cache; 
$32$ KB, $8$-way associative L1 cache;
$2 \times 6$ MB, $8$-way associative L2 cache; and
$8$ MB L3 cache shared by the four cores.
The maximum memory bandwidth of the system is $25.6$ GB/s, 
the on-chip latency is 65 cycles, and the latency between processors is $106$ cycles. 

On Nehalem, the quest for good performance is addressed not only via caching,
that is, exploitation of  spatial and temporal locality, but also through various 
advanced architectural  features. The advanced features include: loop stream detection 
(recognizing instructions belonging to loops and avoiding branch prediction in this context);   
a new MESIF cache coherency protocol that reduces coherency control traffic; 
two branch-target buffers for improved branch prediction;  and out-of-order execution. 
In MESIF, the states are Modified, Exclusive, Shared, Invalid, and a  new Forward state;  
only one instance of a cache line can be in the Forward state, and it is responsible for 
responding to read requests for the cache line, while the  cache lines  in the 
Shared state  remain silent, reducing coherency traffic.
For further architectural details on the Nehalem, see, for example, the paper \cite{Nehalem1}. 

\subsection{The Sun Niagara~2}
\label{sec:niagara}

\begin{figure}
\centering
\includegraphics[scale=0.35]{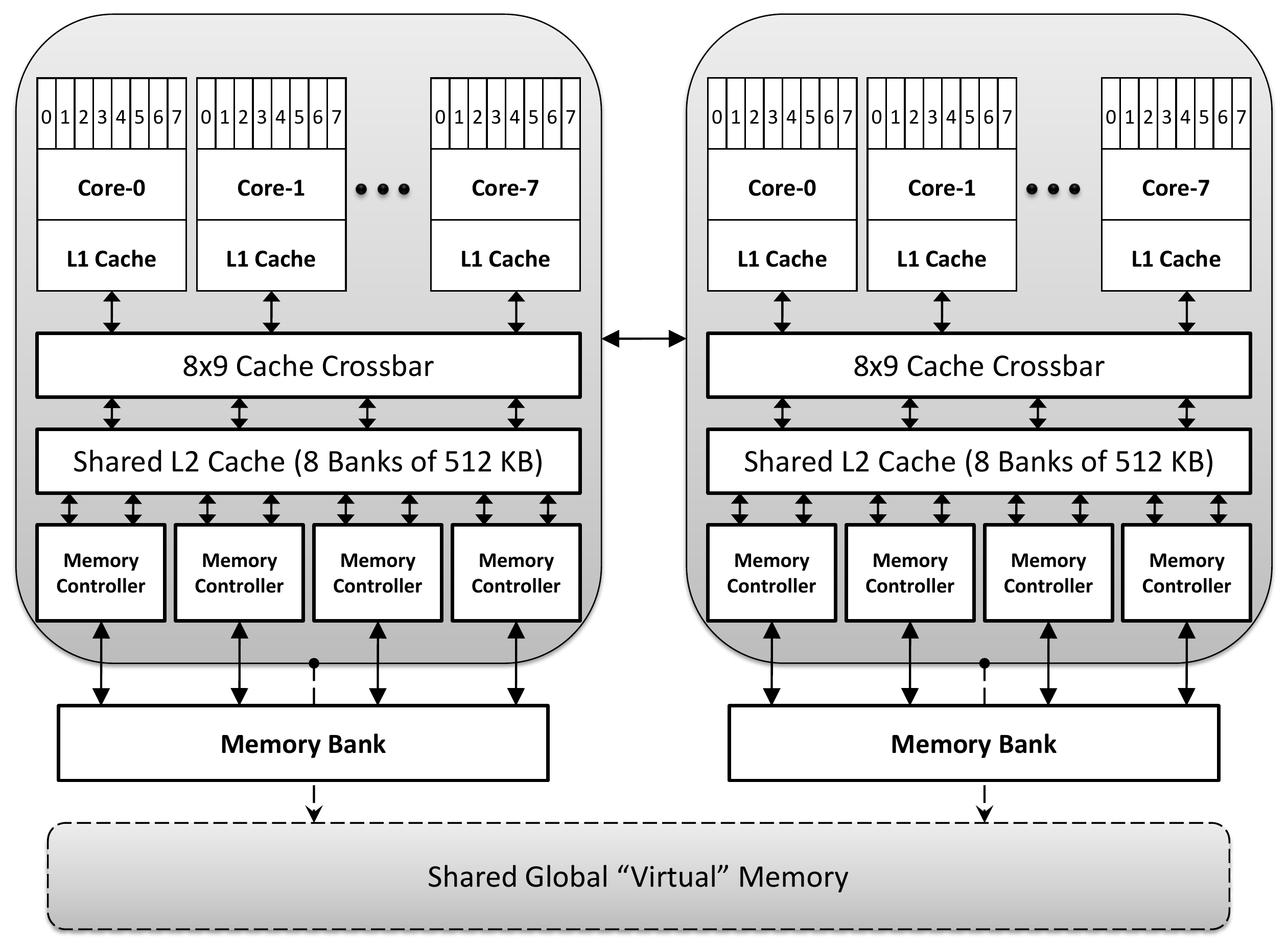}
\caption{Block diagram of the Sun UltraSparc T2 (Niagara~2).} 
\label{fig:niagara}
\end{figure}

The Sun UltraSPARC T2 platform, Niagara~2, has two $8$-core sockets 
with each core supporting $8$ hardware threads.  
Figure~\ref{fig:niagara} shows a block diagram. 
Similar to the Nehalem, multithreading on the Niagara~2  is {\em simultaneous}.
The system has a shallow instruction pipeline (in contrast to the XMT), and each core has
two integer execution units, a load/store unit, and a floating-point unit. The pipeline is capable of 
issuing two instructions per cycle, either from the same thread or from different threads.
Threads are divided into two groups of four, and one instruction from each group may be
selected based on the {\em least-recently fetched} policy on ready threads. 
The clock frequency of the processor  is $1.165$ GHz.

The total size of the memory system is $32$ GB.
Each core has an $8$ KB, $4$-way associative L1 cache for data
and a $16$ KB, $8$-way associative I1 cache for instructions. 
Unlike Nehalem, where L2 cache is private, Niagara~2 has a shared, $16$-way
associative L2 cache of size $4$ MB .
The cache is arranged in $8$ banks and is shared using a crossbar switch between CPUs. 
The latency is $129$ cycles for local memory accesses and $193$ cycles for remote memory accesses. 
The peak memory bandwidth is $50$ GB/s for reads and $26$ GB/s for writes. 
See \cite{Shah-ASCC07} for additional details.

\subsection{The Cray XMT}
\label{sec:xmt}

\begin{figure}
\centering
\includegraphics[scale=0.35]{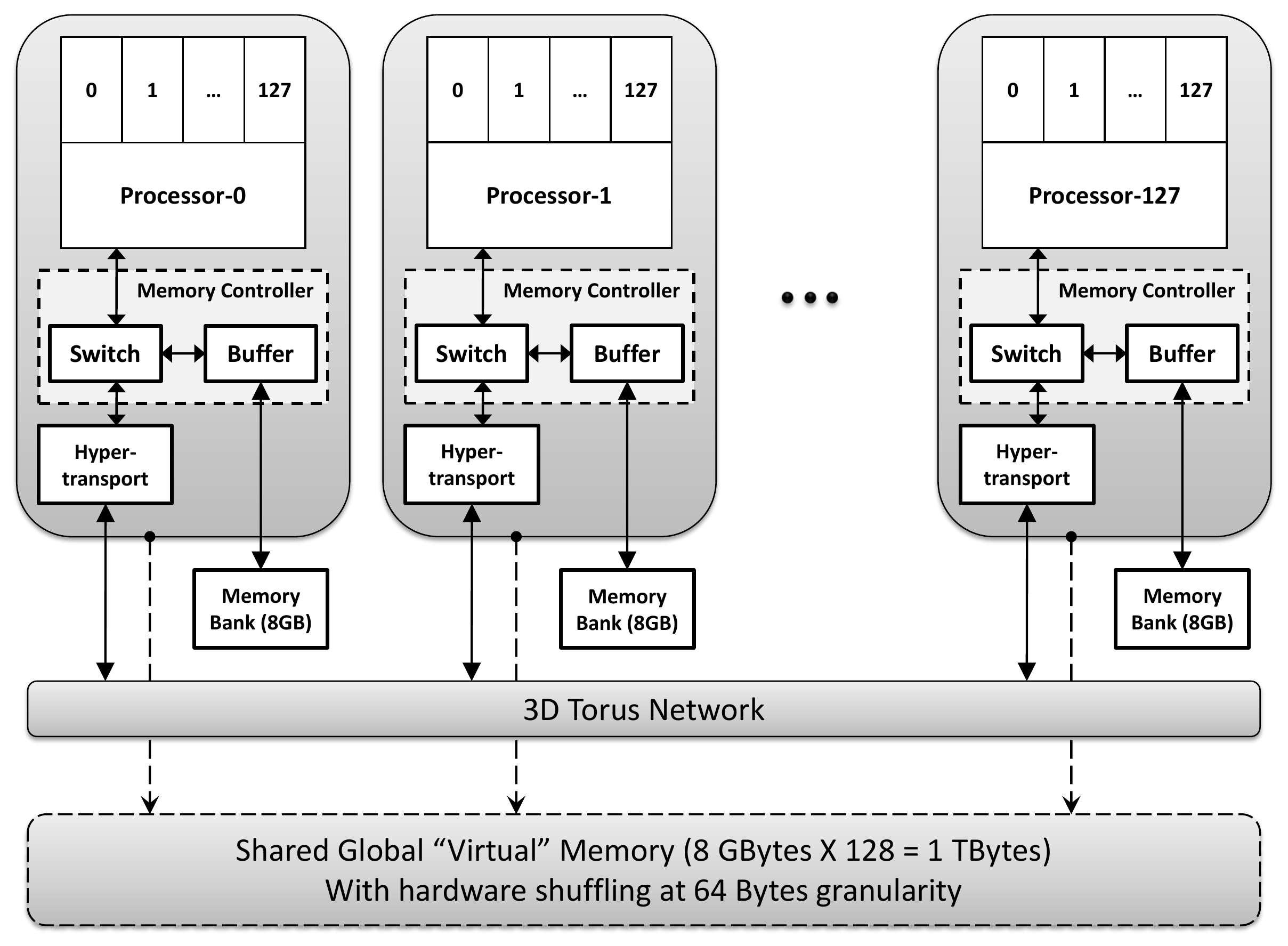}
\caption{Block diagram of the Cray XMT.}
\label{fig:xmt}
\end{figure}

The Cray XMT platform used in this study is comprised of $128$ Cray Threadstorm (MTA-2) 
processors interconnected via a $3$D torus.
Figure~\ref{fig:xmt} shows  a block diagram of the platform from a programmer's point of view.
Each processor has $128$ hardware thread-streams and one instruction pipeline. 
Each thread-stream is equipped with $32$ general purpose registers, 
$8$ target registers, a status word, and a program counter.
Consequently, each processor can maintain up to $128$ separate software threads.  
In every cycle, a processor context switches among threads with ready instructions in a 
fair manner choosing one of the threads to issue its next instruction.  
In other words, the multithreading variant on the XMT, in contrast to Nehalem and Niagara~2, 
is {\em interleaved}.
A processor stalls only if no thread has a ready instruction.
There are three functional  units, M, A, and C  for executing the instructions. 
On every cycle, the M-unit can issue a read or write operation, 
the A-unit can execute a fused multiply-add, and the C-unit can execute either a control or 
an add operation. Instruction execution on the XMT is deeply pipelined, with a $21$-stage pipeline.
The clock frequency is $500$ MHz.

\paragraph{Memory system}
The memory system of the  XMT is cache-less, and is globally accessible by all processors.
The system supports a shared memory programming model.
The global address space is built from physically distributed memory modules of $8$ GB per processor, yielding a total memory size of $1$ TB.
Further, the address space is built using a hardware {\em hashing} mechanism 
that maps the data randomly to the memory modules,  
in blocks of size $64$ bytes,  to  minimize conflicts and hot-spots. 
All memory words are $8$ bytes wide and the memory is byte-addressable.
The worst-case memory latency is approximately $1,000$ cycles, and the average 
value is around $600$ cycles.  
The sustainable minimum remote memory reference rate is $0.52$ GB/s 
per processor for random reads and $0.26$ GB/s per processor for random writes.  
Memory access rates scale with the number of processors.

\paragraph{Fine-grained synchronization}
Associated with every word are several additional bits---a full/empty bit, a pointer forwarding bit,
and two trap bits. These are used for fine-grained  synchronization.
The XMT provides many efficient extended memory semantic operations for manipulating 
the full/empty bits in one clock cycle.
Listed below are examples of functions effecting such semantic operations---many of 
these functions are used in the XMT-specific coloring algorithms we have developed:
\begin{description}
\item[\texttt{purge}] sets the full/empty bit to empty and assigns a zero value 
to the memory location of the associated word. 
\item[\texttt{readff}] reads a memory location only when the full/empty bit is full and leaves the
bit full when the read completes.
\item[\texttt{readfe}] reads a memory location only when the full/empty bit is full and leaves
the bit empty when the read completes.
\item[\texttt{writeef}] writes to a variable only if the full/empty bit is empty and leaves
the bit full when the write completes.
\end{description}
If any extended memory semantic operation finds the full/empty bit in the wrong state, 
the operation waits.
The request returns to the issuing processor and is placed in a queue to be retried.
Note that the instruction is issued only once, and while the instruction is being retried, the processor continues to issue instructions from other threads.
See \cite{John1,Nieplocha07} for further details about the XMT platform.

\subsection{Quick comparison}
\label{sec:comparison}

\begin{table}
{\footnotesize
\centering
\begin{tabular}{l|l|l|l}
\hline \hline
      & {\em Nehalem} & {\em Niagara~2} & {\em Cray XMT} \\    
 \hline \hline        
{\em Clock} (GHz) & 2.53 & 1.165 & 0.5 \\ \hline
{\em Sockets} & 2 & 2 & 128 \\
{\em Cores/socket} & 4 & 8 & -  \\
{\em Threads/core} & 2 & 8 & 128 \\
{\em Threads total} & 16 & 128 & 16,384 \\ \hline
{\em Multithreading} & simultaneous & simultaneous & interleaved \\ \hline
{\em Memory} (GB) & 24 & 32 & 1,024 \\ \hline
{\em Cache} & L1/L2, shared L3 & L1, shared L2 & cache-less, flat \\ \hline
{\em Bandwidth} & max: & peak: &
remote:  \\
& 25.6 GB/s & 50 GB/s (read), & 0.52 GB/s/proc (read),  \\
& & 26 GB/s (write) & 
0.26 GB/s/proc (write) \\ \hline
{\em Latency} &  65 cycles (on-chip), & 129 cycles (local), &
600 cycles (average), \\
&  106 cycles (b/n proc) & 193 cycles (remote) & 1,000 cycles (worst-case) \\
\hline \hline
\end{tabular}

\caption{Overview of architectural features of the platforms Nehalem, Niagara~2 and XMT.}
\label{t:arch-overview}
}
\end{table}   

Table~\ref{t:arch-overview} provides a quick comparative summary of some of the key architectural 
features of the three platforms discussed in Sections~\ref{sec:nehalem} to \ref{sec:xmt}.
Note that the bandwidth and latency numbers listed in Table~\ref{t:arch-overview} are {\em peak values}.
With quick experiments on the Nehalem and Niagara 2 systems using the benchmarking tool
{\em libmicro}, the bandwidth and latency numbers we observed were roughly about a factor
of two worse than those listed in Table~\ref{t:arch-overview}.

\section {The algorithms}
\label{sec:algorithms}

We describe in this section the multithreaded distance-1 coloring algorithms  
we developed for architectures such as those discussed in Section~\ref{sec:platforms}. 
We begin the section with a quick review of the problem, 
the underlying serial algorithm we use to solve it, and prior work on parallelization. 

\subsection{Background}
\label{sec:problem}

\paragraph{The problem}
Graph coloring problems come in several variations \cite{GMP05}.
The variant considered in this paper is {\em distance-1 coloring}, an assignment of  
positive integers (called colors)  to vertices such that  adjacent vertices 
receive different colors.
The objective of the associated  problem is to minimize the number of colors used.
The problem has long been known to be NP-hard to solve optimally.
Recently, it has been shown that, for all $\epsilon > 0$, 
the problem remains NP-hard to {\em approximate} to within $n^{1-\epsilon}$,
where $n$ is the number of vertices in the graph \cite{Zuckerman}.  

\paragraph{An effective algorithm}
\label{sec:sequential}

\begin{algorithm}[t]
\small
\caption{A sequential greedy coloring algorithm.}
\label{algorithm.greedy}
\begin{algorithmic}[1]
\Procedure{Greedy}{$G(V,E)$}
  \State Initialize data structures
  \For{{\bf each} $v \in V$}
      \For{{\bf each} $w \in \mathit{adj}(v)$} 
      \State \textsf{forbiddenColors}[\textsf{color}[$w$]] $\leftarrow$ $v$ \label{l:forbid}
      \Comment{{\footnotesize mark color of $w$ as forbidden to $v$}}
      \EndFor
      \State $c$  $\leftarrow$ $\min\{i>0: \textsf{forbiddenColors}[i] \neq v\}$   \label{l:scolor}  \Comment{{\footnotesize {\em smallest} permissible color}}
      \State \textsf{color}[$v]$ $\leftarrow$ $c$ \label{l:assign}
   \EndFor
\EndProcedure
\end{algorithmic}
\end{algorithm}

Despite these pessimistic theoretical results,
for many graphs or classes of graphs that occur in practice, solutions that are provably optimal
or near-optimal can be obtained using a {\em greedy} (aka sequential) algorithm.
The greedy algorithm runs through the set of vertices in some {\em order}, at each step 
assigning a vertex the {\em smallest} permissible color. 
We give a formal presentation of an {\em efficient} formulation of the greedy algorithm
in procedure \textsc{Greedy} (Algorithm~\ref{algorithm.greedy}).
The formal presentation is needed  since the procedure forms the foundation for the 
parallel algorithms presented  in this paper.

In Algorithm~\ref{algorithm.greedy}, and in other algorithms specified later in this paper, 
$\mathit{adj}(v)$ denotes the set of vertices adjacent to the vertex $v$,
\textsf{color} is a vertex-indexed array that stores the color of each vertex,
and \textsf{forbiddenColors} is a color-indexed array used to mark the colors that are 
impermissible to a particular vertex. 
The array \textsf{color} is initialized at the beginning of the procedure with 
each entry \textsf{color}[$w$] set to zero, to indicate that vertex $w$
is not yet colored.
Each entry of the array \textsf{forbiddenColors} is
initialized {\em once} at the beginning of the procedure
with some value $a \notin V$.
By the end of the inner for-loop in Algorithm~\ref{algorithm.greedy},
all of the colors that are impermissible to the vertex $v$ are recorded in the
array \textsf{forbiddenColors}.
In Line \ref{l:scolor}, the array \textsf{forbiddenColors} is scanned from 
left to right in search of the lowest positive index $i$ at which a value different from $v$,
the vertex being colored, is encountered; this index corresponds to the smallest permissible color
$c$ to the vertex $v$. The color assignment is done in Line \ref{l:assign}.

Note that since the colors impermissible to the vertex $v$ are marked in the array 
\textsf{forbiddenColors} using $v$ as a label, rather than say a boolean flag,  
the array \textsf{forbiddenColors} does not need to be re-initialized in every iteration of 
the loop over the vertex set $V$. Further, the search for color in Line~\ref{l:scolor} terminates
after at most $d(v) + 1$ attempts, where $d(v) = |\mathit{adj}(v)|$ is the degree of vertex $v$.
Therefore, the work done while coloring a vertex $v$ is proportional to 
its degree, independent of the size of the array \textsf{forbiddenColors}.
Thus the time complexity of \textsc{Greedy} is $O(|V| + |E|)$, 
or simply $O(|E|)$ if the graph is connected.
It can also be seen that the number of colors used by the algorithm 
is never larger---and often significantly smaller---than  $\Delta + 1$, 
where $\Delta$ is the maximum degree in the graph.
With good vertex {\em ordering} techniques, the 
number of colors used by \textsc{Greedy} is in fact often near-optimal for
practical graph structures and always bounded by $B + 1$, where $B \leq \Delta$ is
the maximum {\em back degree}---number of already colored neighbors of a vertex---in 
the vertex ordering used by \textsc{Greedy} \cite{colpack-acm}.
A consequence of the fact that the number of colors used by \textsc{Greedy} is 
bounded by $\Delta + 1$ is that the array \textsf{forbiddenColors} need not be of 
size larger than that bound.

\paragraph{Parallelizing the greedy algorithm}

Because of its sequential nature, \textsc{Greedy} is 
challenging to parallelize, when one also wants to keep the number of colors used close
to the serial case. A number of approaches have been 
investigated  in the past to tackle this issue.  One class of the investigated approaches 
relies on iteratively finding a {\em maximal independent set} of vertices in a progressively
shrinking graph and coloring the vertices in the independent set in parallel. 
In many of the methods in this class, the independent set
itself is computed in parallel using some variant of Luby's algorithm \cite{Luby86}.
An example of a method developed along this direction is the work of
Jones and Plassmann \cite{JP93-v2}. Finocchi, Panconesi, and Silvestri~\cite{FPS02}
also follow this direction, but enhance their algorithm in many other ways as well.

Gebremedhin and Manne \cite{GM00} proposed {\em speculation} as an alternative
strategy for coping with the inherent sequentiality of the greedy algorithm.
The idea here is to abandon the requirement that vertices that are colored concurrently
form an independent set, and instead color as many vertices as possible concurrently,
tentatively tolerating potential conflicts, and detect and resolve conflicts afterwards.
A basic shared-memory algorithm based on this strategy is given in \cite{GM00}.
The main steps of the algorithm are to distribute the vertex set equally among the available processors, 
let each processor speculatively color its vertices using information about
already colored vertices (Phase 1), detect eventual conflicts in parallel (Phase 2), 
and finally re-color vertices involved in conflicts sequentially (Phase 3). 

Bozda\u{g} et al.~\cite{BGMBC08} extended the algorithm in \cite{GM00} in a variety of ways 
to make it suitable for and well-performing on distributed memory architectures. 
One of the extensions is replacing the sequential re-coloring phase with 
a parallel iterative procedure.
Many of the other extensions are driven by performance needs in a distributed-memory setting:
the graph needs to be {\em partitioned} among processors in a manner that minimizes
communication cost and maximizes concurrency;
the speculative coloring phase is better when organized in a {\em bulk synchronous} fashion
where computation and communication are coarse-grained; etc.
In addition, they investigate a variety of techniques for choosing colors 
for vertices and relative ordering of
{\em interior} and {\em boundary} vertices, a distinction that arises
due to the partitioning among processors. 
When all the ingredients are put together, it was shown that the 
approach based on speculation and iteration outperforms independent set-based approaches. 
The framework of \cite{BGMBC08} is used for the design of a distributed-memory parallel
algorithm for distance-2 coloring in \cite{BCGMBO:sisc}.

\subsection{Iterative parallel coloring algorithm}
\label{sec:iterative}

In this paper, we adapt the distributed-memory iterative parallel coloring algorithm 
developed in \cite{BGMBC08} to the context of multithreaded, shared-memory platforms.
In the distributed-memory setting, the parallel coloring algorithm relies on graph partitioning 
to distribute and statically map the vertices and coloring tasks to the processors, 
whereas in the multithreaded case the vertex coloring tasks are scheduled 
on processors from a pool of threads created with appropriate programming constructs.

\begin{algorithm}[t]
\small
\caption{An iterative parallel greedy coloring algorithm.} 
\label{algorithm.iterativeparallel}
\begin{algorithmic}[1]
\Procedure{Iterative}{$G(V,E)$}
\State Initialize data structures 
 \State $U \leftarrow V$ \Comment{{\footnotesize $U$ is the current set of vertices to be colored}}
 \While{$U \neq \emptyset$}
      \For{{\bf each} $v\in U$ in {\tt parallel}}  \label{l:phase1} \Comment{{\footnotesize Phase 1: tentative coloring}}
        \For{{\bf each} $w \in \mathit{adj}(v)$}
          \State \textsf{forbiddenColors}[\textsf{color}[$w$]] $\leftarrow$ $v$  \Comment{{\footnotesize thread-private}}
        \EndFor
        \State $c$  $\leftarrow$ $\min\{i>0: \textsf{forbiddenColors}[i] \neq v\}$ 
        \State \textsf{color}[$v$]  $\leftarrow$ c 
     \EndFor   
    \State $R \leftarrow  \emptyset$ \Comment{{\footnotesize $R$ is a set of vertices to be re-colored}}    
     \For{ {\bf each} $v \in U$ in {\tt parallel}} \label{l:phase2}  \Comment{{\footnotesize Phase 2: conflict detection}}
        \For{ {\bf each} $w \in \mathit{adj}(v)$}
          \If{\textsf{color}[$v$] = \textsf{color}[$w$] {\bf and}
           $v > w$} \label{l:conflict}
            \State $R \leftarrow R \cup \{v\}$
          \EndIf
         \EndFor
      \EndFor
     \State $U \leftarrow R$
\EndWhile
\EndProcedure
\end{algorithmic}
\end{algorithm}

The algorithm we use here, formally outlined in Algorithm~\ref{algorithm.iterativeparallel},
proceeds in rounds. Each round has two phases, 
a {\em tentative coloring} phase and a {\em conflict detection} phase, and each phase is 
executed in parallel  over a relevant set of vertices (see lines ~\ref{l:phase1} and \ref{l:phase2}). 
In our implementation, the loops in lines ~\ref{l:phase1} and 
\ref{l:phase2} are parallelized using  the OpenMP directive \texttt{\#pragma omp parallel for}, 
and those parallel loops can  be scheduled in any manner.
The tentative coloring (first) phase is essentially the same as Algorithm~\ref{algorithm.greedy}, 
except that it is concurrently run by multiple threads;
in Algorithm~\ref{algorithm.iterativeparallel}, the array \textsf{forbiddenColors} is 
{\em private} to each thread.
In the conflict-detection (second) phase, each  thread examines a relevant subset
of vertices that are colored in the current round for consistency and identifies a set of vertices
that needs to be re-colored in the next round to resolve any detected conflicts--- a conflict is
said to have occurred when two adjacent vertices get the same color.
Given a pair of adjacent vertices involved in a conflict, 
it suffices to recolor only one of the two to resolve the conflict. 
In Algorithm~\ref{algorithm.iterativeparallel}, as can be seen in Line~\ref{l:conflict}, 
the vertex with the higher {\em index} value is chosen to be re-colored in the event of a conflict. 
Algorithm~\ref{algorithm.iterativeparallel} terminates when no more 
vertices to be re-colored are left. 

Assuming that the number of rounds required by Algorithm~\ref{algorithm.iterativeparallel} is
sufficiently small, the overall work performed by the algorithm is linear in the input size.
This assumption is realistic and is supported empirically by the experimental results  
to be presented in Section~\ref{sec:results}.

\subsection{Cray XMT-specific algorithms}
\label{sec:dataflow}

We implemented Algorithm~\ref{algorithm.iterativeparallel} on the Cray XMT by replacing  the
OpenMP directive  \texttt{\#pragma omp parallel for} used to parallelize lines~\ref{l:phase1}  and \ref{l:phase2} 
with the XMT directive \texttt{\#pragma mta assert parallel}.
Typically, parallel algorithms written for conventional processors are too coarse-grained 
to effectively use the massive multithreading capability provided by the XMT; 
hence, we were motivated to explore the potential of an alternative algorithm that employs {\em fine-grained}
synchronization of threads and avoids the occurrence of conflicts (hence the need for re-coloring)
in greedy coloring.

\subsubsection{A dataflow algorithm}

\begin{algorithm}[t]
\small
\caption{A dataflow algorithm for coloring.} 
\label{algorithm.dataflow.coloring}
\begin{algorithmic}[1]
\Procedure{Dataflow}{$G(V,E)$}
  \For{{\bf each} $v \in V$ in {\tt parallel}} \label{l:purge}
      \State \texttt{purge}(\textsf{color}[$v$]) \Comment{{\footnotesize Sets full/empty bit to empty and value to zero}}
  \EndFor
  \For{{\bf each} $v \in V$ in {\tt parallel}} \label{l:dcolor}
      \For{{\bf each} $w \in \mathit{adj}(v)$ where $w<v$}
              \State $c\_w$ $\leftarrow$ \texttt{readff}(\textsf{color}[$w$])       
              \Comment{{\footnotesize Wait until full/empty bit becomes full}} \label{dataflow:readff-line}
             \State \textsf{forbiddenColors}[$c\_w$] $\leftarrow$ $v$ \Comment{{\footnotesize thread-private}}
      \EndFor
      \State $c$  $\leftarrow$ $\min\{i>0: \textsf{forbiddenColors}[i] \neq v\}$  
      \State \texttt{writeef}(\textsf{color}[$v$], $c$) 
          \Comment{{\footnotesize Write color and set full/empty bit to full}}
  \EndFor
\EndProcedure
\end{algorithmic}
\end{algorithm}

Using the extended memory semantic operations of the XMT functions discussed in 
Section~\ref{sec:xmt} for fine-grained synchronization of threads, 
we developed a dataflow algorithm for coloring, which is formally 
outlined in \textsc{Dataflow} (Algorithm~\ref{algorithm.dataflow.coloring}).
The algorithm consists of two parallel loops.
The first loop {\em purges} the array used to store the color of each vertex.
The second loop runs over all vertices with thread $T_v$ responsible for coloring vertex $v$.
It has two steps.
First, the thread $T_v$ reads the color chosen by the neighbors of $v$ with {\em higher precedence}.
We use the {\em indices} of the vertices to define the precedence
order such that vertices with smaller indices will have higher precedence.
Second, having read the colors, the thread $T_v$ chooses the smallest permissible color and 
assigns it to the vertex $v$.
Because a \texttt{readff} is used to read the colors of neighbors of $v$ and 
a \texttt{writeef} to write the color of $v$, the algorithm is race-free.
Leaving the XMT-specific mechanism in which concurrent execution
is made possible aside, note that this algorithm is conceptually the same as the 
Jones-Plassmann algorithm \cite{JP93-v2}.

\subsubsection{A recursive dataflow algorithm}

As written, the algorithm \textsc{Dataflow} may deadlock on a system with fewer compute resources
than vertices, since only a subset of the threads can start; the others need to wait for resources to
become available. If all active threads read vertices whose threads are waiting, deadlock occurs.

The deadlock can be eliminated in one of three ways: 1) preemption, 2) ordering, or 3) recursion.
Preempting blocked threads is expensive, and is a poor choice when 
high-performance and scalability are important goals.
Ordering the threads to ensure progress may not be possible in all cases, 
and in many cases, computing a correct order is expensive.
Using recursion is a simple and often efficient way to eliminate deadlock in dataflow algorithms 
executing on systems with limited compute resources.

We outline the deadlock-free recursive variant of the dataflow coloring
algorithm in procedure \textsc{DataflowRecursive} (Algorithm~\ref{algorithm.dataflow.recursive}).
The function \textsc{ProcessVertex} (Algorithm~\ref{algorithm.process.vertex}) 
is now responsible for coloring vertex $v$.
Its code is similar to the body of the second loop of
\textsc{Dataflow} (Algorithm~\ref{algorithm.dataflow.coloring}).
Before reading a neighbor's color, the procedure checks whether or not the neighbor is processed.
If the neighbor is processed, the procedure issues a \texttt{readff} to read its color; if the neighbor
is not processed, the procedure processes the neighbor itself.
The recursive algorithm stores the state of each vertex in a new array named \textsf{state}; a value of
zero indicates that a vertex is unprocessed. 
The algorithm also utilizes the XMT intrinsic function \texttt{int\_fetch\_add} to manipulate
the state of each vertex.
The function is an atomic operation that increments the value of a given memory location 
with a given constant and returns the {\em original} value.  
Note that the procedure \textsc{ProcessVertex} is called only once per vertex---when its 
state is zero. Hence the overall work performed by the algorithm is linear in the input size. 

\begin{algorithm}[t]
\small
\caption{Recursive dataflow algorithm for coloring.}
\label{algorithm.dataflow.recursive}
\begin{algorithmic}[1]
\Procedure{DataflowRecursive}{$G(V,E)$}
  \For{{\bf each} $v \in V$ in {\tt parallel}}
      \State \texttt{purge}(\textsf{color}[$v$]) 
      \Comment{{\footnotesize Sets full/empty bit to empty and value to zero}}
      \State \textsf{state}[$v$]$\gets 0$
  \EndFor
  \For{{\bf each} $v \in V$ in {\tt parallel}}
      \State CurState $\gets$ \texttt{int\_fetch\_add}(\textsf{state}[$v$], 1) 
      \Comment{{\footnotesize returns 0 if first time}}
      \If{CurState $= 0$}
           \State \textsc{ProcessVertex}$(v)$
       \EndIf
  \EndFor
\EndProcedure
\end{algorithmic}
\end{algorithm}

\begin{algorithm}[t]
\small
\caption{A routine called by \textsc{DataflowRecursive} (Algorithm~\ref{algorithm.dataflow.recursive}).}
\label{algorithm.process.vertex}
\begin{algorithmic}[1]
\Procedure{ProcessVertex}{$v$}
      \For{{\bf each} $w \in \mathit{adj}(v)$ where $w<v$}
             \State CurState $\gets$ \texttt{int\_fetch\_add}(\textsf{state}[$w$], 1)
             \If{CurState $= 0$}
                \State \textsc{ProcessVertex}$(w)$ \Comment{{\footnotesize Recursive call}}
             \EndIf
             \State $c\_w$ $\leftarrow$ \texttt{readff}(\textsf{color}[$w$])
                 \Comment{{\footnotesize Wait until full/empty bit becomes full}} \label{processvertex-readff-line} 
          \State \textsf{forbiddenColors}[$c\_w$] $\leftarrow$ $v$ \Comment{{\footnotesize thread-private}}
   \EndFor
   \State $c$  $\leftarrow$ $\min\{i>0: \textsf{forbiddenColors}[i] \neq v\}$  
   \State \texttt{writeef}(\textsf{color}[$v$], $c$) \Comment{{\footnotesize Write color and set full/empty bit to full}}
\EndProcedure
\end{algorithmic}
\end{algorithm}

\section{Test Graphs}
\label{sec:graphs}

We studied the performance of algorithms \textsc{Iterative} and \textsc{DataflowRecursive}
on the three platforms discussed in Section~\ref{sec:platforms}
using a set of {\em synthetically generated} graphs.
The test graphs are designed to  represent a wide spectrum of input types. 
We briefly discuss in this section the algorithm used to 
generate the test graphs and the key structural properties 
(relevant to the performance of the coloring algorithms) 
of the generated graphs.

\subsection{The graph generator}

The test graphs were created using the R-MAT graph generator~\cite{Chakrabarti}.
Let the graph to be generated consist of $|V|$ vertices and $|E|$ edges.
The R-MAT algorithm works by recursively subdividing the adjacency matrix of the graph to 
be generated (a $|V|$ by $|V|$ matrix) into four quadrants $(1,1)$, $(1,2)$, $(2,1)$, and
$(2,2)$, and distributing the $|E|$ edges within the quadrants with specified probabilities.
The distribution is determined by four non-negative parameters ($a$, $b$, $c$, $d$) 
whose sum equals one. Initially every entry of the adjacency matrix is zero (no edges added).
The algorithm places an edge in the matrix by choosing one of the four quadrants
$(1,1)$, $(1,2)$, $(2,1)$, or, $(2,2)$ with probabilities $a$, $b$, $c$, or, $d$, respectively.
The chosen quadrant is then subdivided into four smaller partitions and the procedure is
repeated until a $1 \times 1$ quadrant is reached, where the entry is incremented 
(that is, the edge is placed).
The algorithm repeats the edge generation process $|E|$ times to create the desired graph
$G=(V,E)$.  

By choosing the four parameters appropriately, graphs of varying characteristics can be generated. 
We generated three graphs of the same size but widely varying structures by 
using the following set of parameters: \\
$(0.25, 0.25, 0.25, 0.25)$; $(0.45, 0.15, 0.15, 0.25)$; $(0.55, 0.15, 0.15, 0.15)$. 

We call the three graphs RMAT-ER, RMAT-G, and RMAT-B, respectively.
(The suffix ER stands for ``Erd\H os-R\'enyi,''
G and B stand for ``good" and ``bad" for reasons that will be apparent shortly.)
Table~\ref{t:prop} provides basic statistics on the structures of the three graphs.
The graphs were generated on the XMT using an implementation of the R-MAT algorithm 
similar to the one  described in~\cite{SSCA}. 
The graphs were then saved on disk for reuse on other platforms. 
Duplicate edges and self loops were removed;
the small variation in the number of edges is due to such removals. 
The sizes of the graphs were chosen so that they would fit on 
all three of the platforms we consider. 
For scalability study on the XMT, however, we generated {\em larger} graphs 
using the same three sets of four parameters; 
we shall encounter these in Section~\ref{sec:results}, Table~\ref{t:LargeScale}.

\begin{table}
\begin{footnotesize}
\centering
\begin{tabular}{|l|l|l|l|r|r|}
\hline
\textbf{Graph} & \textbf{No. } & \textbf{No.} &  \textbf{Avg.} &
\textbf{Max.} & \textbf{Variance}  \\ 
& \textbf{Vertices} & \textbf{Edges} &  \textbf{Deg.} &
\textbf{Deg.} & \\ \hline \hline
RMAT-ER		&  16,777,216 &    134,217,654 &  16 & 42     &  16.01    \\ \hline
RMAT-G	        &  16,777,216 &    134,181,095 &  16 & 1,278  &  415.72   \\ \hline
RMAT-B	        &  16,777,216 &    133,658,229 &  16 & 38,143 &  8,085.64 \\ \hline
\end{tabular}
\caption{Structural properties of the test graphs.}
\label{t:prop}
\end{footnotesize}
\end{table}

\subsection{Characteristics of the test graphs}

The test graphs are {\em designed} to represent instances 
posing varying levels of difficulty for the performance of the multithreaded coloring algorithms. 
The three graphs in Table~\ref{t:prop} are nearly identical in size, 
implying that the serial work involved in the algorithms is the same, 
but they vary widely in degree distribution of the vertices and
density of local subgraphs. Both of these have implications on available concurrency in algorithms \textsc{Iterative} and \textsc{DataflowRecursive}: 
large-degree vertices and dense subgraphs would increase the likelihood for conflicts to occur 
in \textsc{Iterative} and would cause more ``serialization" in \textsc{DataflowRecursive}, 
thereby impacting parallel performance.

\begin{figure}
\centering
\subfigure[RMAT-ER]{\includegraphics[width=0.48\textwidth]{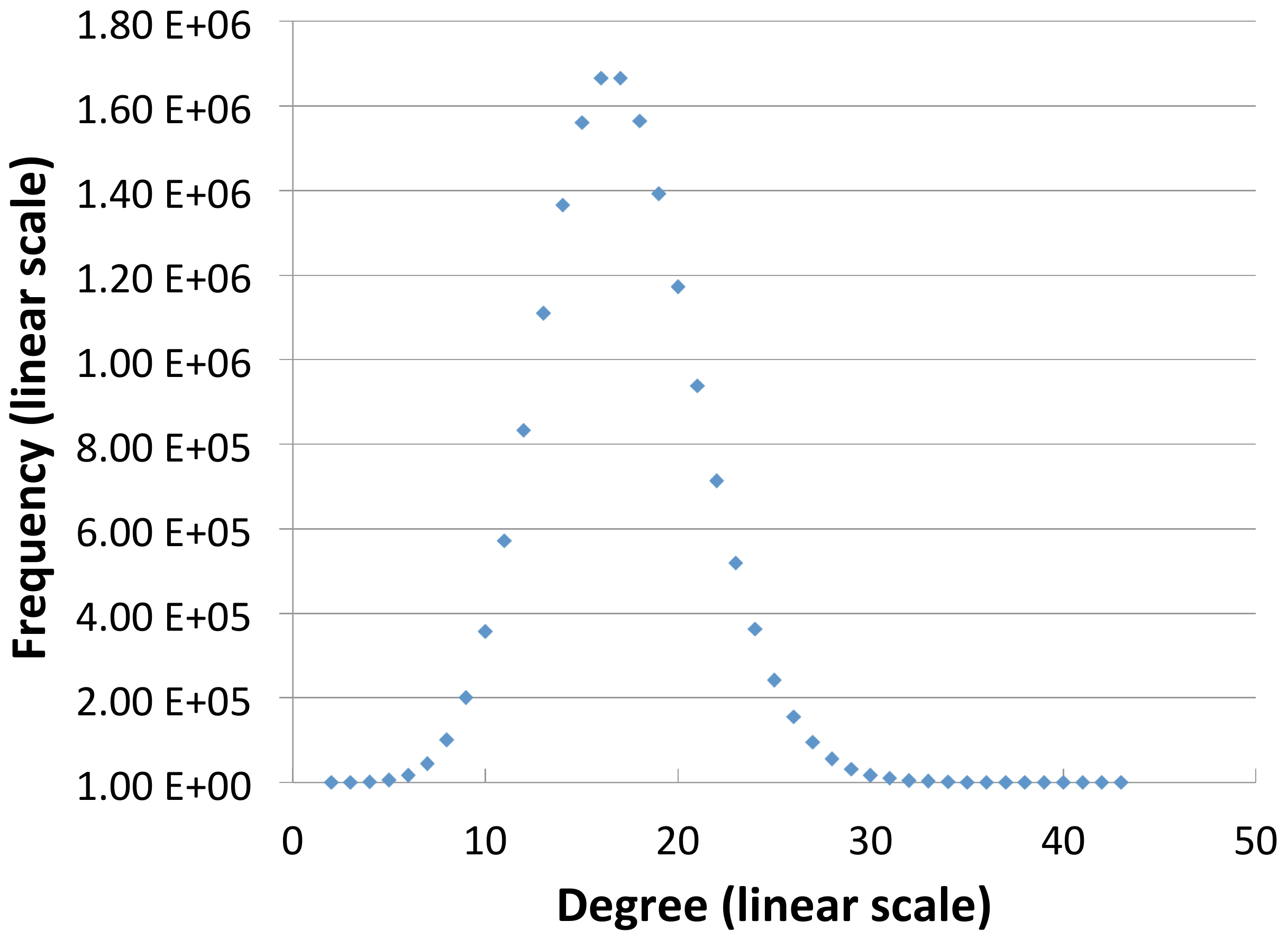}\label{f:degDist-ER}}
\subfigure[RMAT-G]{\includegraphics[width=0.48\textwidth]{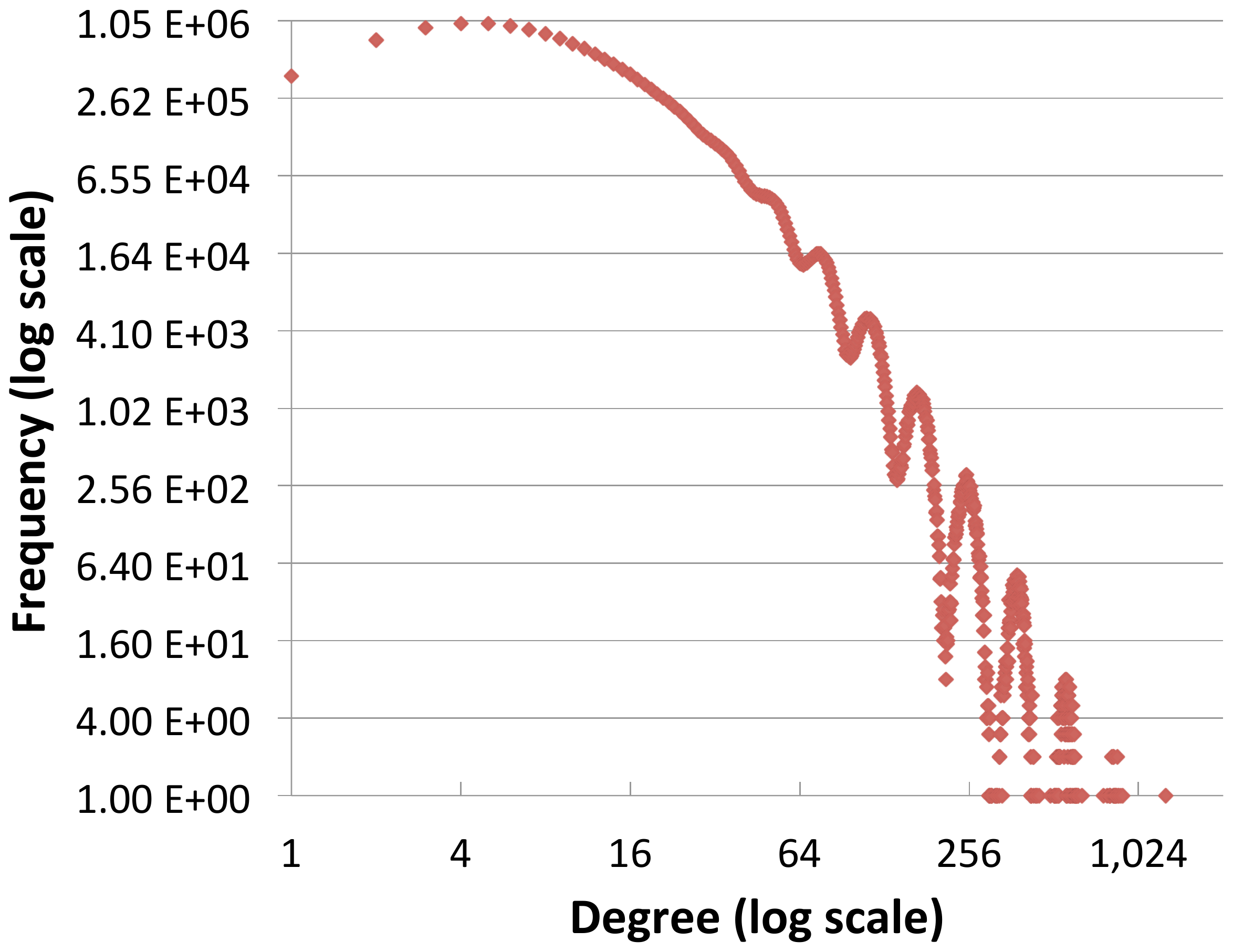}\label{f:degDist-G}}
\subfigure[RMAT-B]{\includegraphics[width=0.48\textwidth]{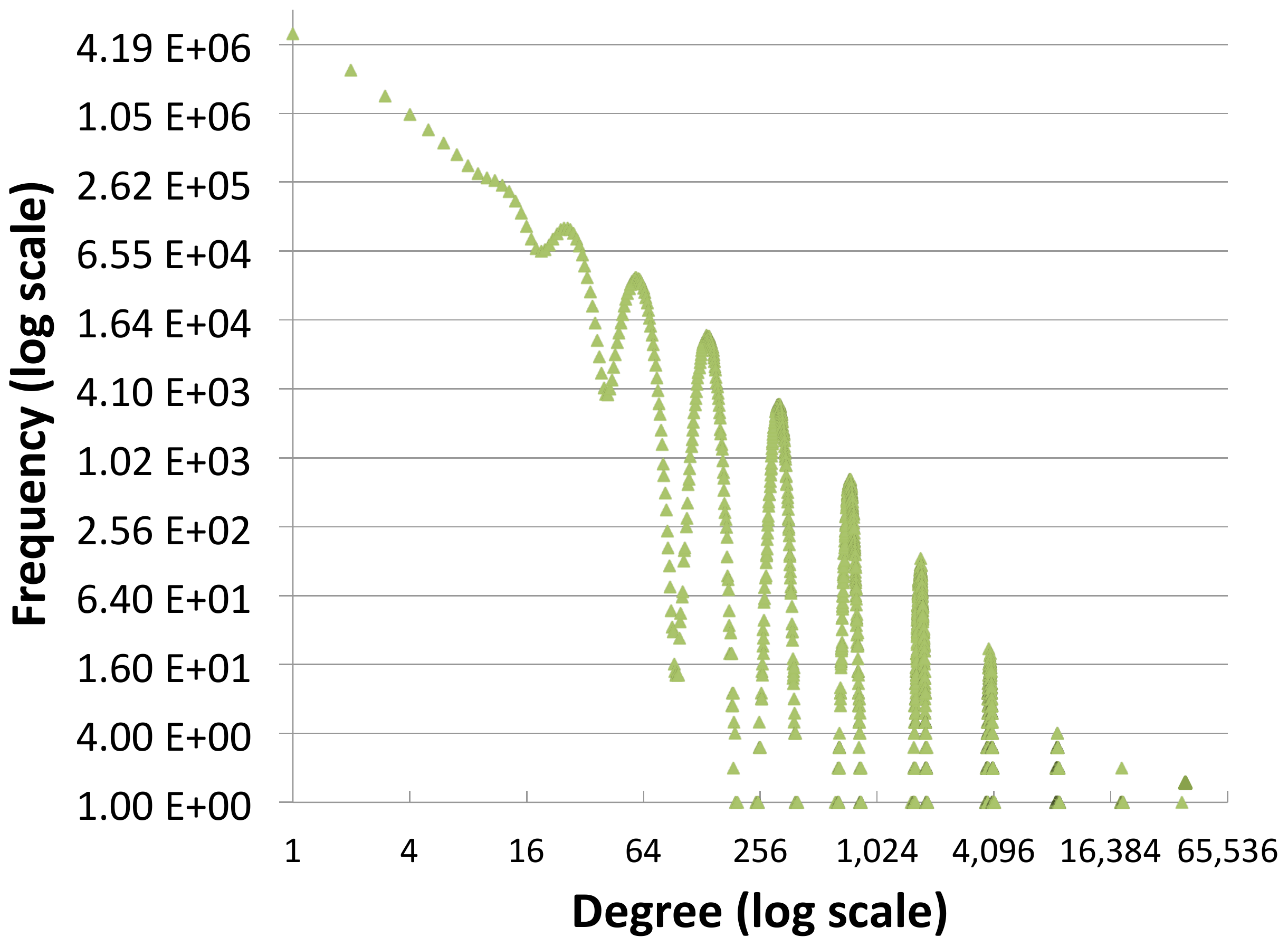}\label{f:degDist-B}}
\subfigure[All three graphs]{\includegraphics[width=0.48\textwidth]{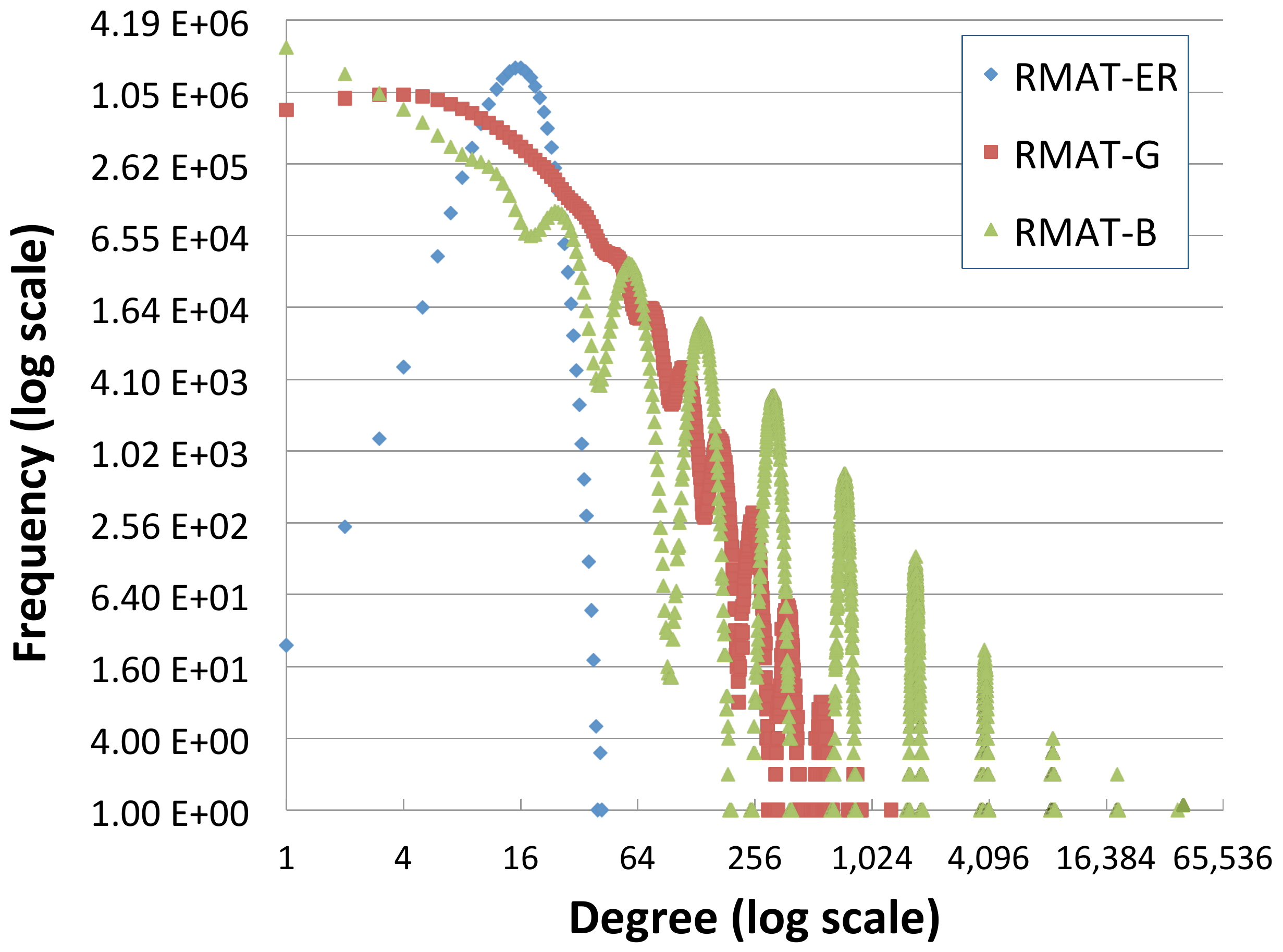}\label{f:degDist-ALL}}
\caption{Degree distribution of the three test graphs listed in Table~\ref{t:prop}.}
\label{f:deg-Dist}
\end{figure}

Figure~\ref{f:deg-Dist} summarizes the vertex degree distributions in the three graphs.
The graph RMAT-ER belongs to the class of Erd\H os-R\'enyi random graphs,
since $a=b=c=d=0.25$. Its degree distribution is expected to be {\em normal}.
The single local maximum observed in the corresponding curve in the degree distribution plot
depicted in Figure~\ref{f:degDist-ER} is consistent with this expectation.
In the graphs RMAT-G and RMAT-B, ``subcommunities'' (dense local subgraphs)  
are expected to form because of the larger values of $a$ and $d$ 
compared to $b$ and $c$; the subcommunities are coupled together 
in accordance with the values of $b$ and $c$.
The multiple local maxima observed in the degree distributions of these two graphs,
shown in Figures~\ref{f:degDist-G} and \ref{f:degDist-B}, 
correlate with the existence of such communities.

Besides degree distribution, the three test graphs  also vary highly
in terms of maximum degrees and degree variances (see Table~\ref{t:prop}).
As an additional measure for the structural variation represented by the three graphs, 
we computed the clustering coefficients of vertices in these graphs
using the tool GraphCT\footnote{Available at http://trac.research.cc.gatech.edu/graphs/wiki/GraphCT}.
The {\em local clustering coefficient} of a vertex $v$ is the ratio between the actual number of edges 
among the neighbors $\mathit{adj}(v)$ of $v$ to the total possible number of edges 
among $\mathit{adj}(v)$ \cite{citeulike:99}. 
The average clustering coefficient in the graph is the sum of all local clustering coefficients 
of vertices divided by the number of vertices.

\begin{figure}
\subfigure[Average clustering coefficient]{
\begin{minipage}[c][0.51\textwidth]{0.35\textwidth}
\centering
\small
\begin{tabular}{l|l}
                  & Avg. CC \\ \hline 
RMAT-ER & $1.00 e-6$ \\
RMAT-G   & $1.20 e-5$ \\
RMAT-B    & $3.43 e-4$ \\ 
\end{tabular}
\label{f:AvgCC}
\end{minipage}
}
\subfigure[Local clustering coefficient distribution]{
\begin{minipage}[c][0.51\textwidth]{0.61\textwidth}
\centering
\includegraphics[width=1.0\textwidth]{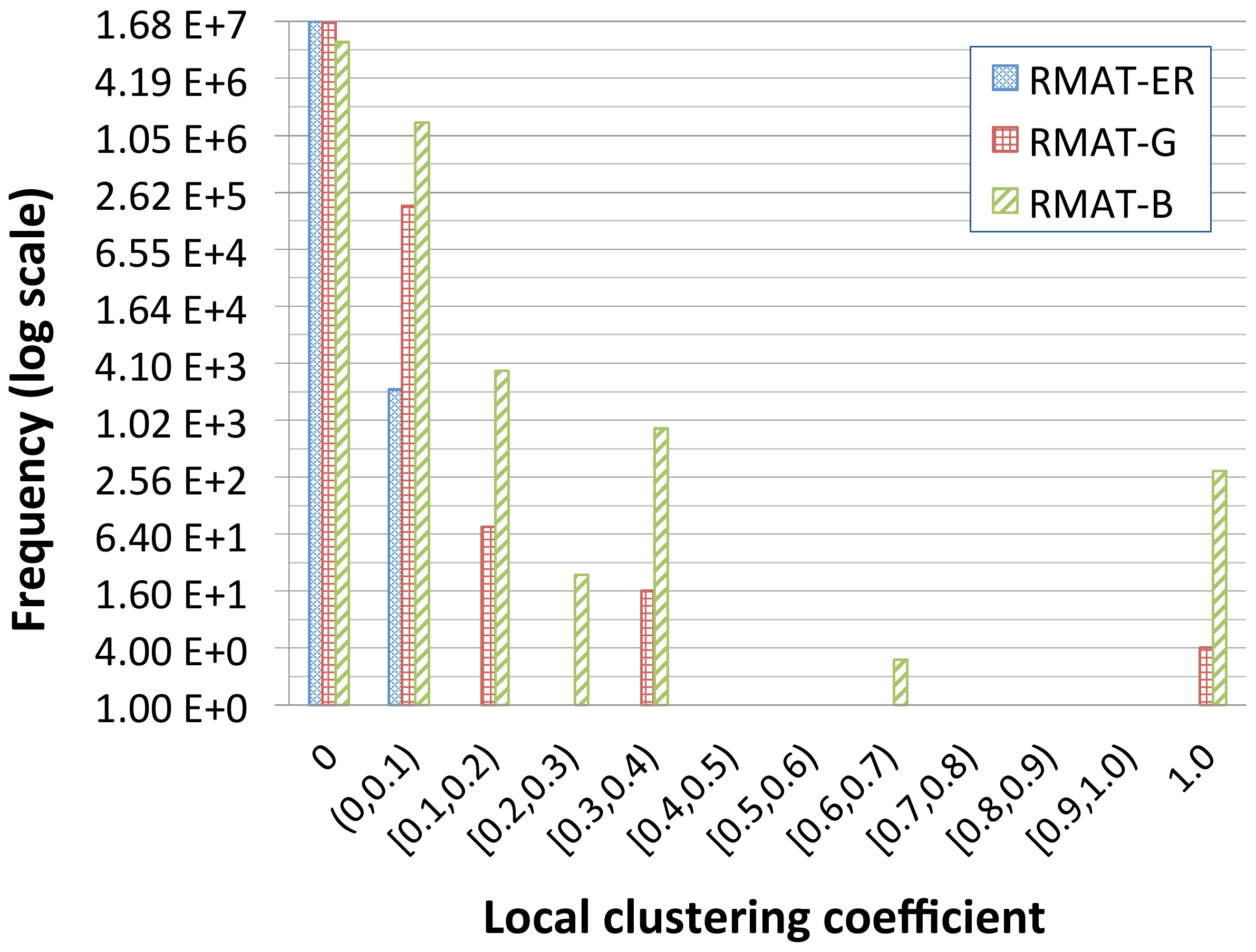}\label{f:CCdist}
\end{minipage}
}
\caption{Clustering coefficient data on the three test graphs listed in Table~\ref{t:prop}.}
\label{f:cc}
\end{figure}

Tabulated in Figure~\ref{f:AvgCC} is the average clustering coefficient in each of the three test graphs. 
The numbers there show orders of magnitude progression as one goes from 
RMAT-ER to RMAT-G to RMAT-B. 
Figure~\ref{f:CCdist} shows the distribution of local clustering coefficients, 
categorized in  blocks of values. For the RMAT-ER graph, the local clustering coefficient
of any vertex is found to be either 0, or between 0 and 0.1, whereas the values are spread
over a wider  range for the RMAT-G graph and even wider range for the RMAT-B graph.
Larger local clustering coefficients in general indicate the presence of relatively 
dense local subgraphs, or subcommunities, as referred to earlier.

\begin{figure}
\centering
\subfigure[RMAT-ER]{\includegraphics[width=0.325\textwidth]{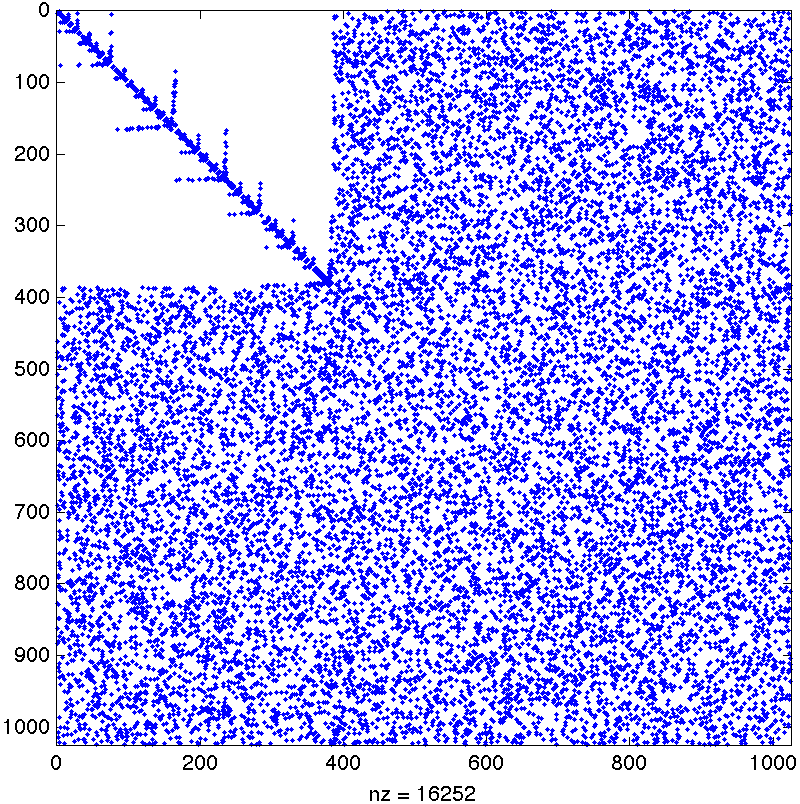}\label{f:ER-amd}}
\subfigure[RMAT-G]{\includegraphics[width=0.325\textwidth]{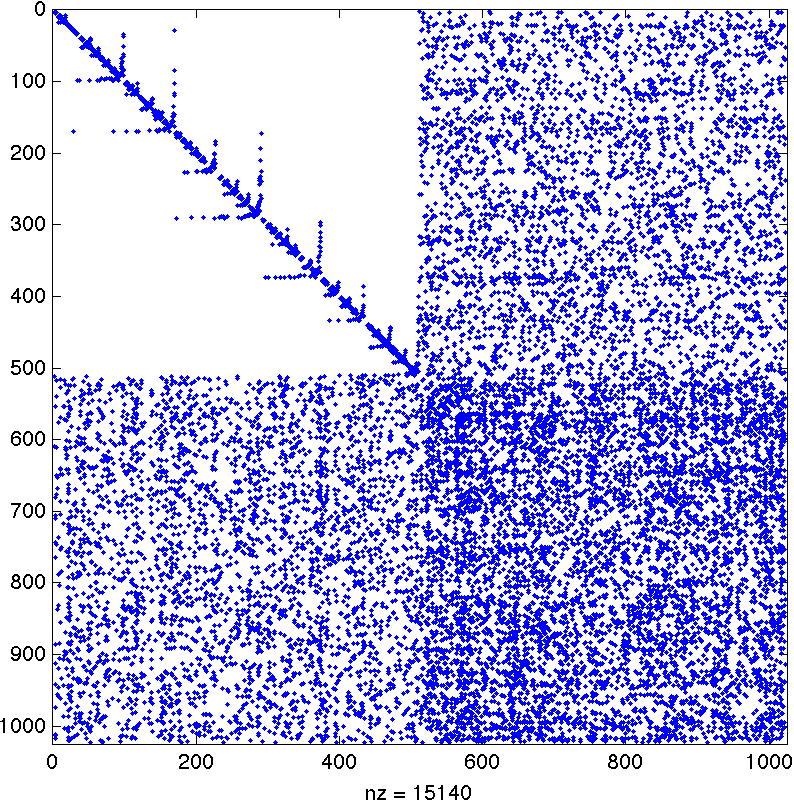}\label{f:good-amd}}
\subfigure[RMAT-B]{\includegraphics[width=0.325\textwidth]{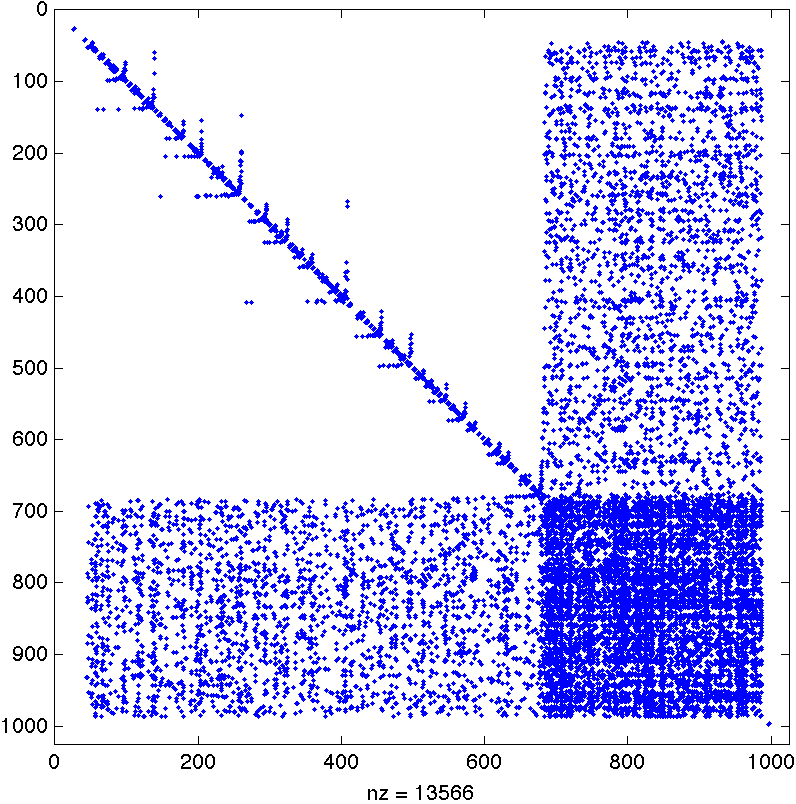}\label{f:bad-amd}}
\caption{{\tt Spy} plots of the adjacency matrices of scaled-down versions of
the three graphs with the columns reordered using Approximate Minimum Degree to aid visual appeal.
Each matrix has $2^{10} = $ 1,024 columns and 16,252 nonzeros. }
\label{f:spyplots-amd}
\end{figure}

Finally, to visualize the difference in the structure represented by the three graphs,
we include in Figure~\ref{f:spyplots-amd} plots generated with
MATLAB's {\tt spy} function of the adjacency matrices of scaled-down versions of the three
test graphs (1,024 vertices and 16,252 edges). 
To aid visual appeal, in the {\tt spy}-plots, the columns have been re-ordered
using the Approximate Minimum Degree (AMD) routine in MATLAB;  loosely speaking, 
AMD orders lower degree vertices (columns) before higher degree vertices 
in a graph model of sparse Gaussian elimination that updates 
the degrees of vertices dynamically \cite{amestoy+}. 

Based on the variation in maximum degree value, degree distribution, and
clustering coefficient values the three test graphs represent, one can see that a 
fairly wide spectrum of input types is covered in the experimental study.

\section{Experimental Results}
\label{sec:results}

With the test graphs discussed in Section~\ref{sec:graphs} serving as inputs, 
we present in this section experimental results on the scalability of and the number of colors 
used  by the algorithms \textsc{Iterative} and \textsc{DataflowRecursive}
when they are run on the three platforms described in Section~\ref{sec:platforms}. 
We begin the section by discussing various matters on experimental setup 
in Sections~\ref{sec:locality} through \ref{sec:binding}.
The scalability results are presented in 
Sections~\ref{sec:scalability-sep}, \ref{sec:scalability-comp}, and \ref{sec:iterative-more}, 
and the results on the number of colors used are presented in Section~\ref{sec:colors}.
The purpose of Section~\ref{sec:scalability-sep} is to discuss the
scalability results on each of the three platforms {\em separately} while that of
Section~\ref{sec:scalability-comp} is to draw comparisons.
Section~\ref{sec:iterative-more} shows results analyzing the performance of 
\textsc{Iterative} in more detail.

\subsection{Locality Not Exploited}
\label{sec:locality}

The R-MAT algorithm tends to generate graphs where most high-degree vertices are of low indices.  
In the tests we run, we {\em randomly} shuffled  vertex indices in order to reduce
the benefits of this artifact  on architectures with caches and to minimize memory-hotspotting.
The vertices were then colored in the resulting numerical order in both algorithms. 

\subsection{Compilers Used}
\label{sec:compilers}
Algorithm \textsc{Iterative} was implemented on the Nehalem and Niagara~2  platforms using OpenMP.
On the Cray XMT, both  \textsc{Iterative} and \textsc{DataflowRecursive} were 
implemented using the XMT programming model.
Table~\ref{t:compilers} summarizes the compilers and flags used on each platform.
\begin{table}
\begin{footnotesize}
\centering
\begin{tabular}{l|ll}
& Compiler & Flag \\ \hline
Nehalem & Intel 11.1 & -fast \\
Niagara~2 & Sun Studio 5.9 & -fast -xopenmp -m64 \\
XMT & Cray Native C/C++  & -par  
\end{tabular}
\caption{Overview of compilers and flags used in the experiments.}
\label{t:compilers}
\end{footnotesize}
\end{table}

\subsection{Thread Scheduling}
\label{sec:scheduling}

In both \textsc{Iterative} and \textsc{DataflowRecursive}, 
there is a degree of freedom in the choice
of thread {\em scheduling policy} for parallel execution.  
Both the OpenMP and the XMT programming models offer directives 
supporting different kinds of policies.
We experimented with  various policies (static,
dynamic, and guided with different block sizes)
on the three platforms to assess the impact of the policies on performance. 
We observed no major difference in performance for the experimental setup 
we have, that is, RMAT-generated graphs with vertex indices randomly shuffled.
Note, however, that for vertex orderings that exhibit
some locality (e.g., some high degree vertices are clustered together) 
scheduling policies could
affect performance due to  cache effects.
In the results reported in this paper, we have used the default {\em static} 
scheduling on all platforms.

\subsection{Thread Binding}
\label{sec:binding}

In the experiments on scalability, we used various platform-specific thread binding mechanisms.    
On the Nehalem, binding  was achieved using a {\em high-level affinity interface},
the environment variable \texttt{KMP\_AFFINITY} from the Intel compiler. 
Recall that the Nehalem has two sockets with four cores per socket, and
two threads per core, amounting to $16$ threads in total. 
The eight threads on the four cores of the first socket are
numbered (procID) as 
$\{(1, 9), (3, 11), (5, 13), (7, 15)\}$, 
where the first pair corresponds to the threads
on the first core, the second pair to the threads on the second core, etc.
Similarly the remaining eight  threads on the second socket are numbered as
 $\{(0, 8), (2, 10), (4, 12), (6, 14)\}$.
With such topology in place, we bound the OpenMP threads to the cores by
assigning the procID to \texttt{KMP\_AFFINITY}. For example, 
to run a job with four OpenMP threads on four cores (one thread per core), we set \\
 \texttt{KMP\_AFFINITY="verbose,proclist=[9,8,15,14]"} 
and \texttt{OMP\_NUM\_THREADS=4}.\\
Notice here that threads from the two sockets, rather than from
only one, are engaged. Note also that the threads are bound to cores that are as far apart as possible.
These seemingly counter-intuitive choices are made to maximize bandwidth.
Recall that the memory banks in the Nehalem system (see Figure~\ref{fig:nehalem}) 
are associated with sockets, and each socket has independent channels to its memory bank. 
Engaging the two sockets would thus increase the effective memory bandwidth available. 
Such a choice could, however, be counter-productive for small problems that fit in a single bank, 
or for cache reuse between threads. 

Niagara~2 provides a mechanism for binding the threads to processors using
\texttt{SUNW\_MP\_PROCBIND}. This was exploited in the experiments by
binding $1$, $2$, $4$, or $8$ threads to each core to get results from $1$ to $128$ threads. 

\subsection{Scalability Results}
\label{sec:scalability-sep}

\subsubsection{Scalability of \textsc{Iterative} on Nehalem and Niagara~2}

\begin{figure}
\centering
\subfigure[RMAT-ER (Nehalem)]{\includegraphics[width=0.47\textwidth]{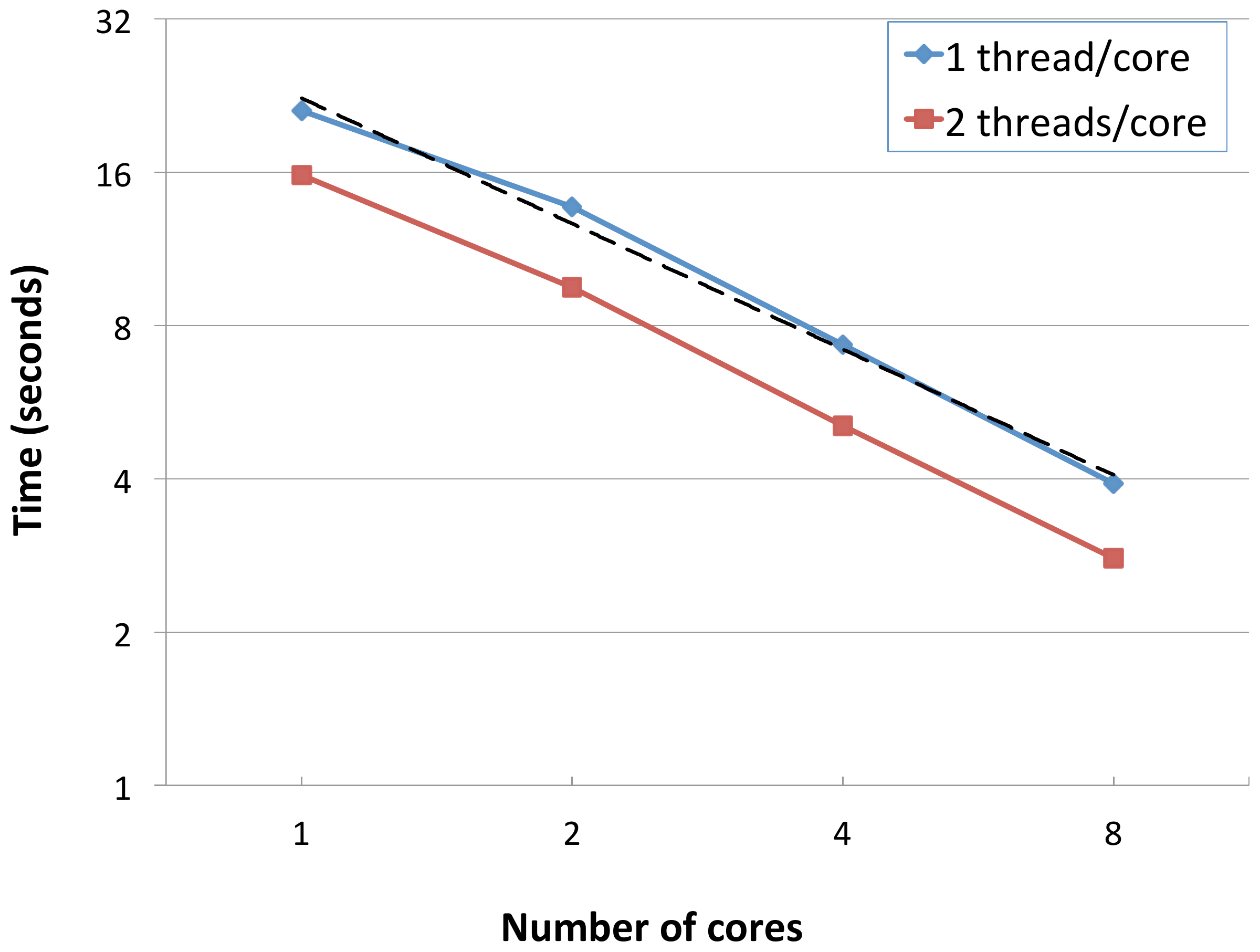}
  \label{f:nehalem-er}}
\subfigure[RMAT-ER (Niagara~2)]{\includegraphics[width=0.47\textwidth]{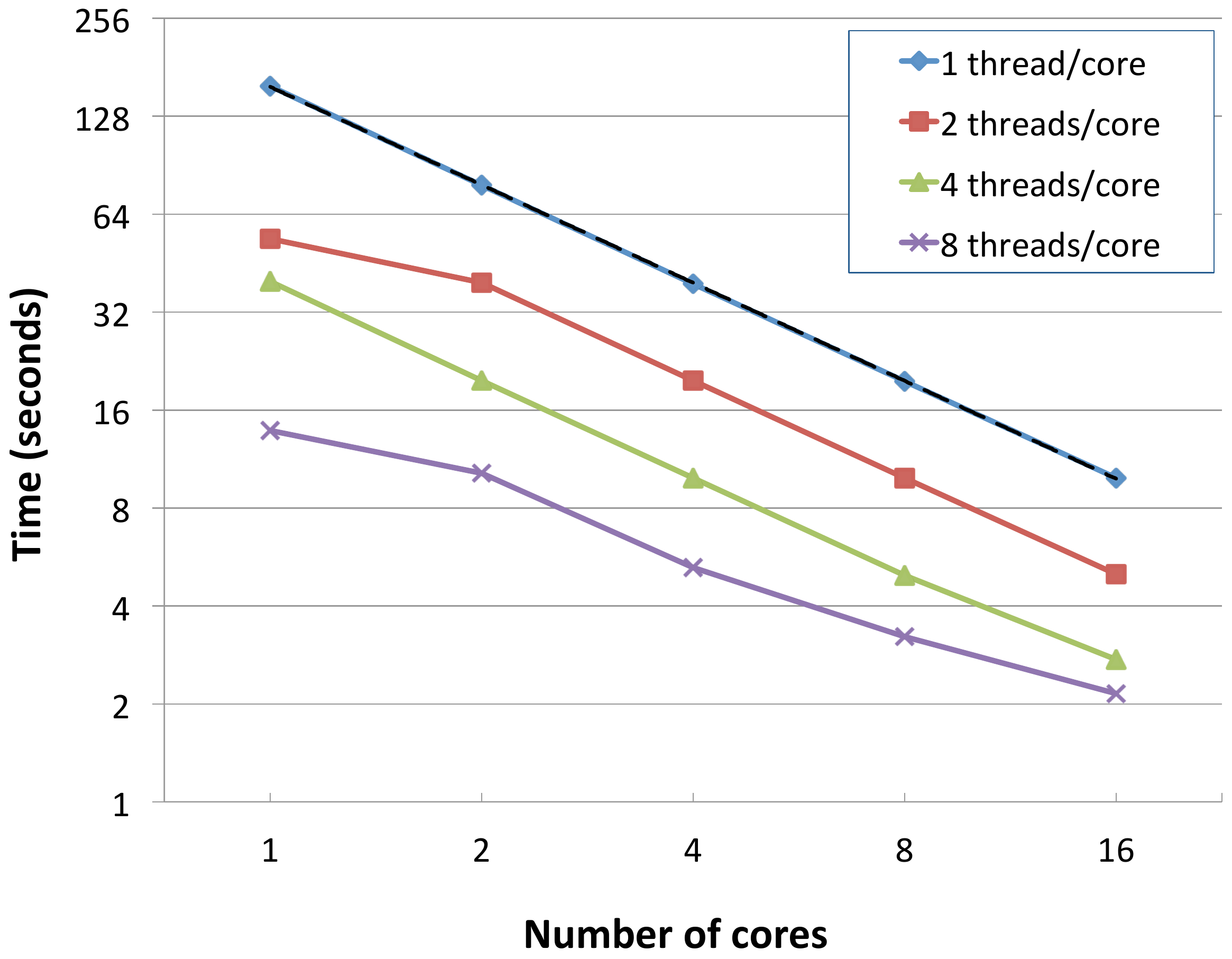}
  \label{f:niagara-er}}
\subfigure[RMAT-G (Nehalem)]{\includegraphics[width=0.47\textwidth]{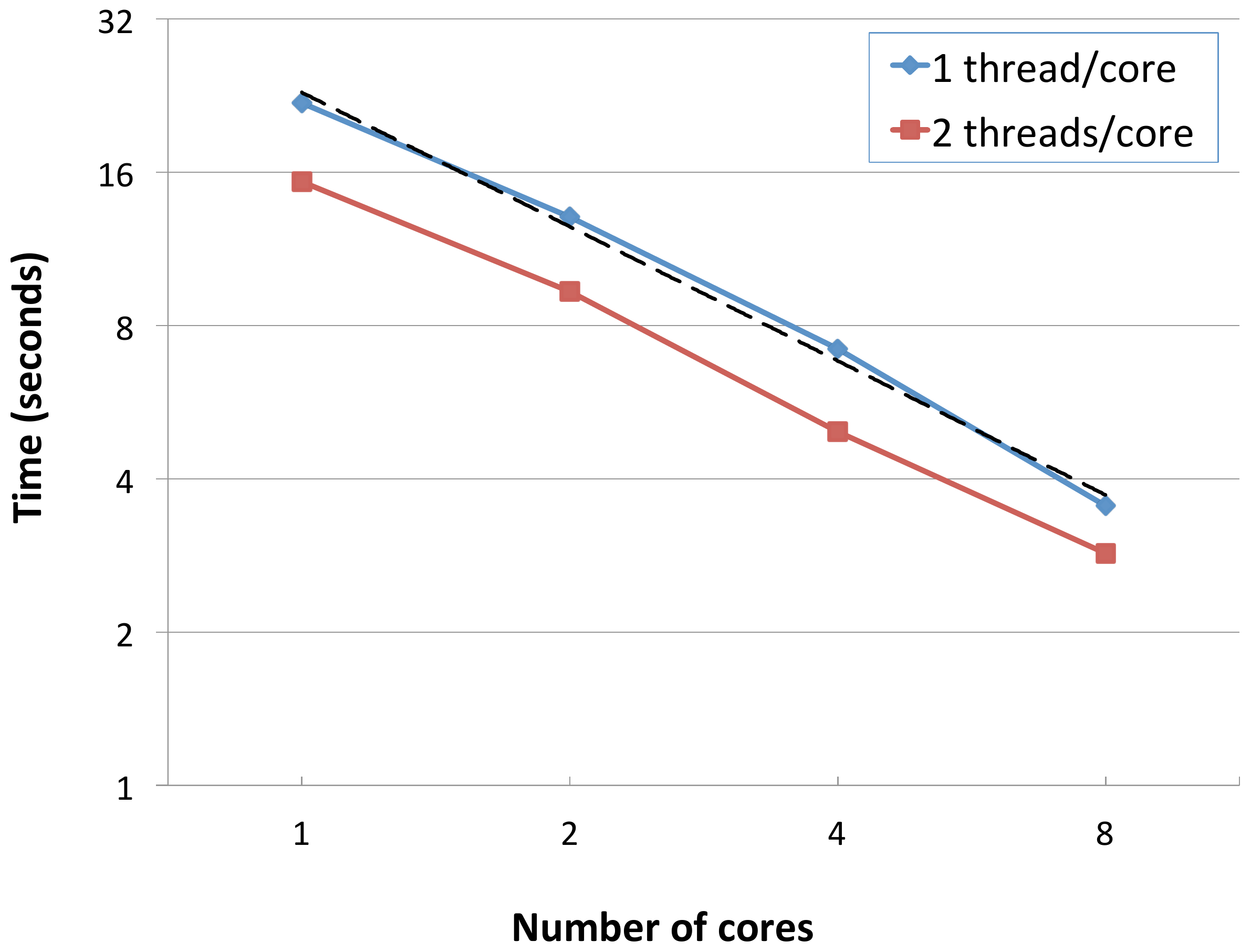}
  \label{f:nehalem-good}}
\subfigure[RMAT-G (Niagara~2)]{\includegraphics[width=0.47\textwidth]{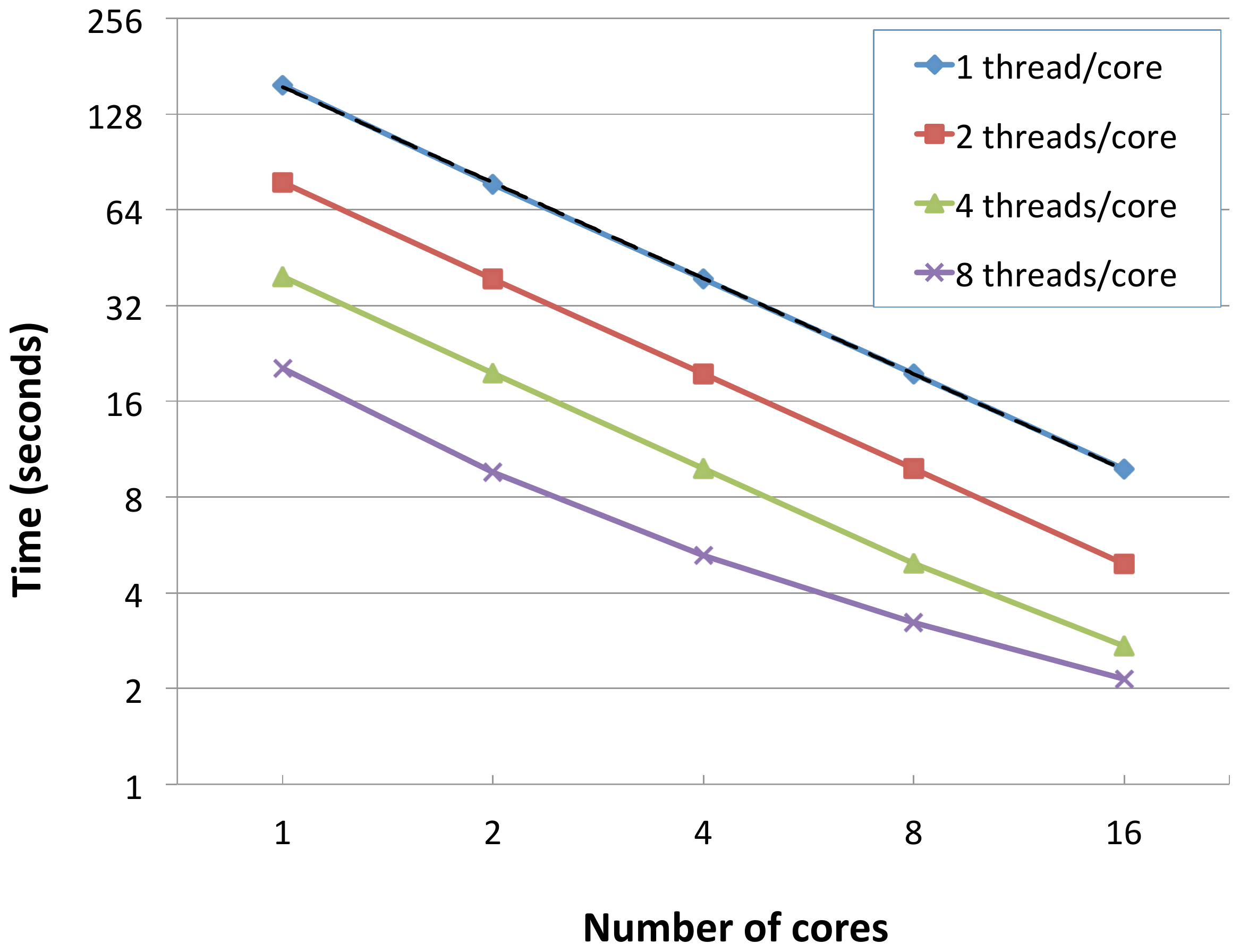}
  \label{f:niagara-good}}
\subfigure[RMAT-B (Nehalem)]{\includegraphics[width=0.47\textwidth]{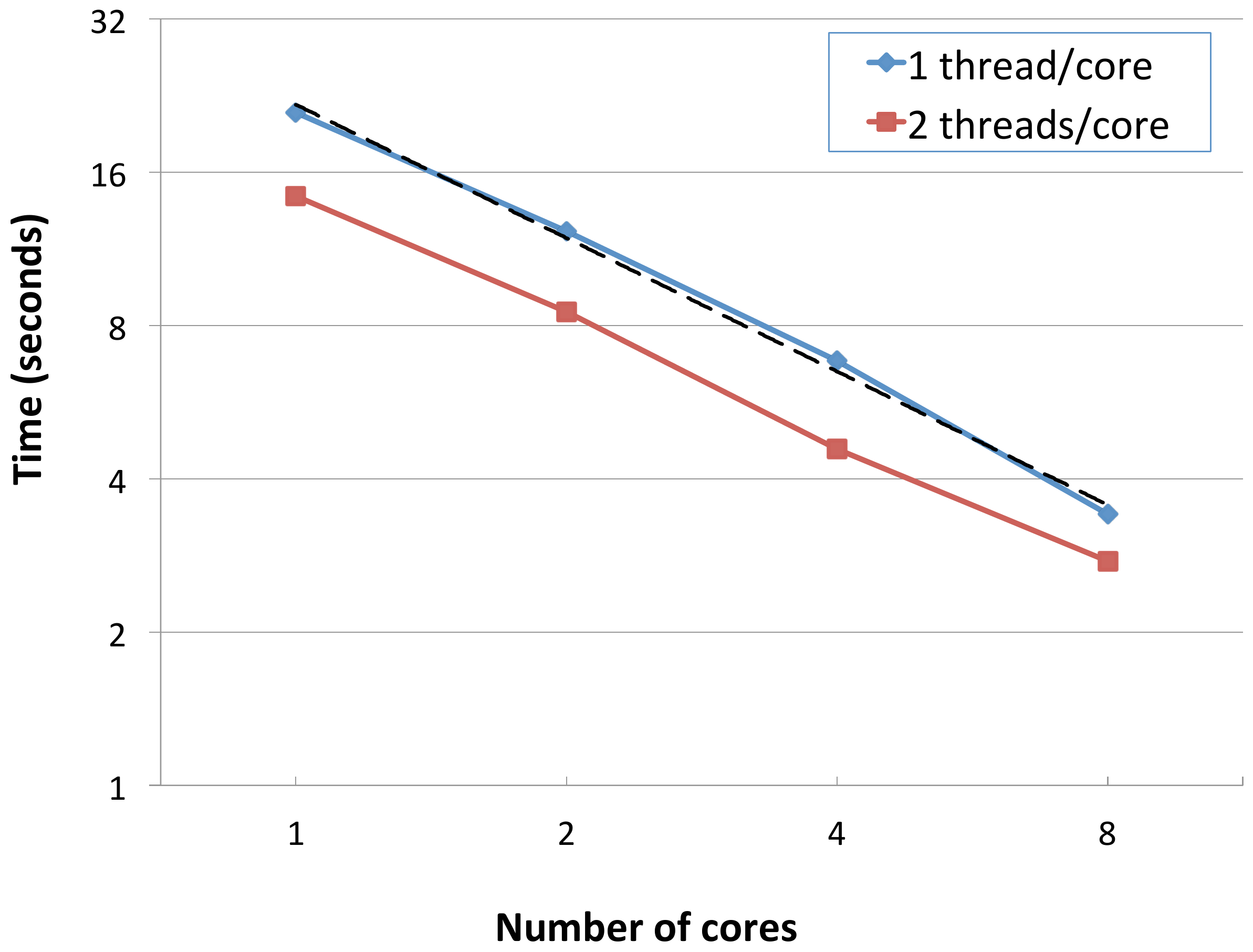}
  \label{f:nehalem-bad}}
\subfigure[RMAT-B (Niagara~2)]{\includegraphics[width=0.47\textwidth]{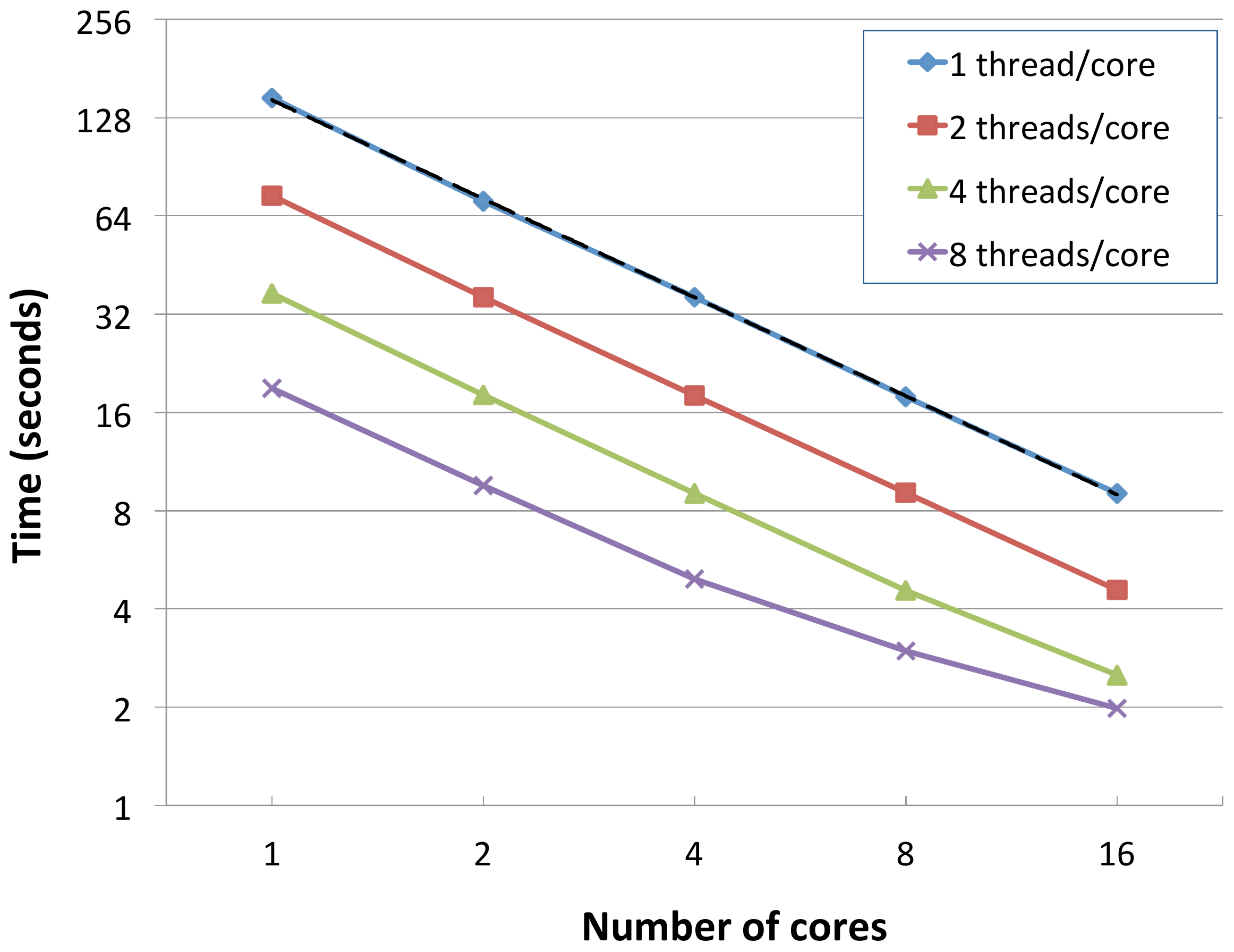}
  \label{f:niagara-bad}}
\caption{\small Strong scaling of the \textsc{Iterative} algorithm on Nehalem and Niagara~2
for the graphs listed in Table~\ref{t:prop}.
On each platform, results for different number of threads per core are shown. 
The dashed-line curve in each chart corresponds to {\em ideal} speedup in the case where 
one thread per core is used. In all charts, both axes are in logarithmic scale.}
\label{f:nehalem}
\label{f:niagara}
\end{figure}

Figure~\ref{f:nehalem} gives a summary of strong scaling results of algorithm \textsc{Iterative} 
on the three test graphs RMAT-ER, RMAT-G, and RMAT-B listed in Table~\ref{t:prop}.
The left column shows results on the Nehalem platform, 
and the right column shows results on the Niagara~2 platform. 
On each platform, different numbers of threads per core were utilized via thread binding.
In the sets of results on both architectures, the curve
represented by the dashed line corresponds to {\em ideal} speedup
in the case where one thread per core is used.
It can be seen that the strong scaling behavior of algorithm \textsc{Iterative} is near-ideal
on both architectures across all three input types considered, when one thread per core is used.

The results in Figure~\ref{f:nehalem} corresponding to the use of multiple threads per core 
on both the Nehalem and Niagara~2 show the benefit of exploiting simultaneous multithreading (SMT).
On the Nehalem, by using two threads per core, instead of one,
we obtained relative speedups ranging  from $1.2$ to $1.5$, depending on the number of cores 
used and the type of input graph considered.
The observed speedup due to SMT is less than the ideal factor of two, and the difference
is more pronounced when the number of cores increases to  eight. 
Note that the speedup is greater for the RMAT-ER graph than for the RMAT-B and RMAT-G graphs.
This is because the memory accesses in the RMAT-ER graphs
pose less performance challenges than the accesses in the RMAT-B and RMAT-G graphs, 
due to the presence of a larger number of dense subgraphs in the latter two. 
The speedup we obtained via the use of SMT on the Nehalem is
consistent with the results reported in \cite{Barker},
where the authors measured the runtime reduction (speedup) obtained by 
employing SMT on the Nehalem on seven application codes. 
The authors found only a 50 percent reduction in the best case, 
and for four of the codes, they observed an increase in runtime (slowdown), 
instead of decrease (speedup). 

The speedup due to SMT we obtained on the Niagara~2 
(the right column in Figure~\ref{f:nehalem}) is more impressive than that on the Nehalem. 
By and large, we observed that the runtime of \textsc{Iterative} using $2T$ threads on $C$ cores is
about the same as the runtime using $T$ threads on $2C$ cores, except when  $T$ is $4$, in which case the relative gain due to SMT is smaller.

\subsubsection{Scalability of \textsc{Iterative} and \textsc{DataflowRecursive}
on the XMT}

\begin{table}
\begin{footnotesize}
\centering
\begin{tabular}{|r|r|r|r|r|r|r|}
\hline
\textbf{Graph}	& \textbf{Scale}	&	\textbf{No. }	&	\textbf{No. }	&	\textbf{Max.}	&	
                   \textbf{Variance}	&	\textbf{\% Isolated}	\\  
& &\textbf{Vertices}	& \textbf{Edges} &	\textbf{Deg.}	& &	\textbf{vertices}	\\ \hline \hline                   
\multirow{4}{*}{RMAT-ER} & 24	&	 16,777,216 	&	 134,217,654 	&	 42 		&	 16.00 		& 0	\\ 
& 25	&	 33,554,432 	&	 268,435,385 	&	 41 		&	 16.00 		& 0   	\\ 
& 26	&	 67,108,864 	&	 536,870,837 	&	 48 		&	 16.00 		& 0   	\\ 
& 27	&	 134,217,728 	&	 1,073,741,753 	&	 43 		&	 16.00 		& 0   	\\ \hline
\multirow{4}{*}{RMAT-G}	& 24	&	 16,777,216 	&	 134,181,095 	&	 1,278 		&	 415.72 	& 2.33 	\\ 
& 25	&	 33,554,432 	&	 268,385,483 	&	 1,489 		&	 441.99 	& 2.56 	\\ 
& 26	&	 67,108,864 	&	 536,803,101 	&	 1,800 		&	 469.43 	& 2.81 	\\ 
& 27	&	 134,217,728 	&	 1,073,650,024 	&	 2,160 		&	 497.88 	& 3.06 	\\ \hline
\multirow{4}{*}{RMAT-B} & 24	&	 16,777,216 	&	 133,658,229 	&	 38,143 	&	 8,085.64 	& 30.81 \\ 
& 25	&	 33,554,432 	&	 267,592,474 	&	 54,974 	&	 9,539.17 	& 32.34 \\ 
& 26	&	 67,108,864 	&	 535,599,280 	&	 77,844 	&	 11,213.79 	& 33.87 \\ 
& 27	&	 134,217,728 	&	 1,071,833,624 	&	 111,702 	&	 13,165.52 	& 35.37 \\ \hline
\end{tabular}
\caption{\small Properties of the larger graphs used in the scalability study on the XMT.}
\label{t:LargeScale}
\end{footnotesize}
\end{table}

{\em Larger test graphs. } 
The Cray XMT provides about $1$ TB of global shared memory and
massive  concurrency ($128 \times 128 = $ 16,384-way interleaved multithreading).
Both capabilities are much larger than what is available on the other two platforms.
To make use of this huge resource and assess scalability, 
we experimented with larger RMAT graphs generated using the same parameters as
in the experiments thus far but with different {\em scales}.
The characteristics of these graphs are summarized in Table~\ref{t:LargeScale}. 
In the table, the number of vertices (third column) is equal to $2^s$, where $s$ is the {\em scale}
shown in the second column.

{\em Thread allocation. }
In contrast to the Niagara~2 and Nehalem, where thread allocation and binding is {\em static}, 
the corresponding tasks on the XMT are {\em dynamic}. 
However, one can place a system request for a maximum number of {\em streams} 
(the XMT term for threads) for a given execution on the XMT. 
In our experiments, we requested $100$ streams,
an empirically determined optimal number for keeping all processors busy. 
The runtime system then decides as to how many will actually get allocated. 
We observed that at the beginning of the execution of the coloring algorithms, 
we do get close to $100$ streams. Then, the number drastically decreases 
and gets to around $5$ towards the end of the execution, when little computational work is left.

\begin{figure}
\centering
\subfigure[RMAT-ER (\textsc{Iterative})]{\includegraphics[width=0.47\textwidth]{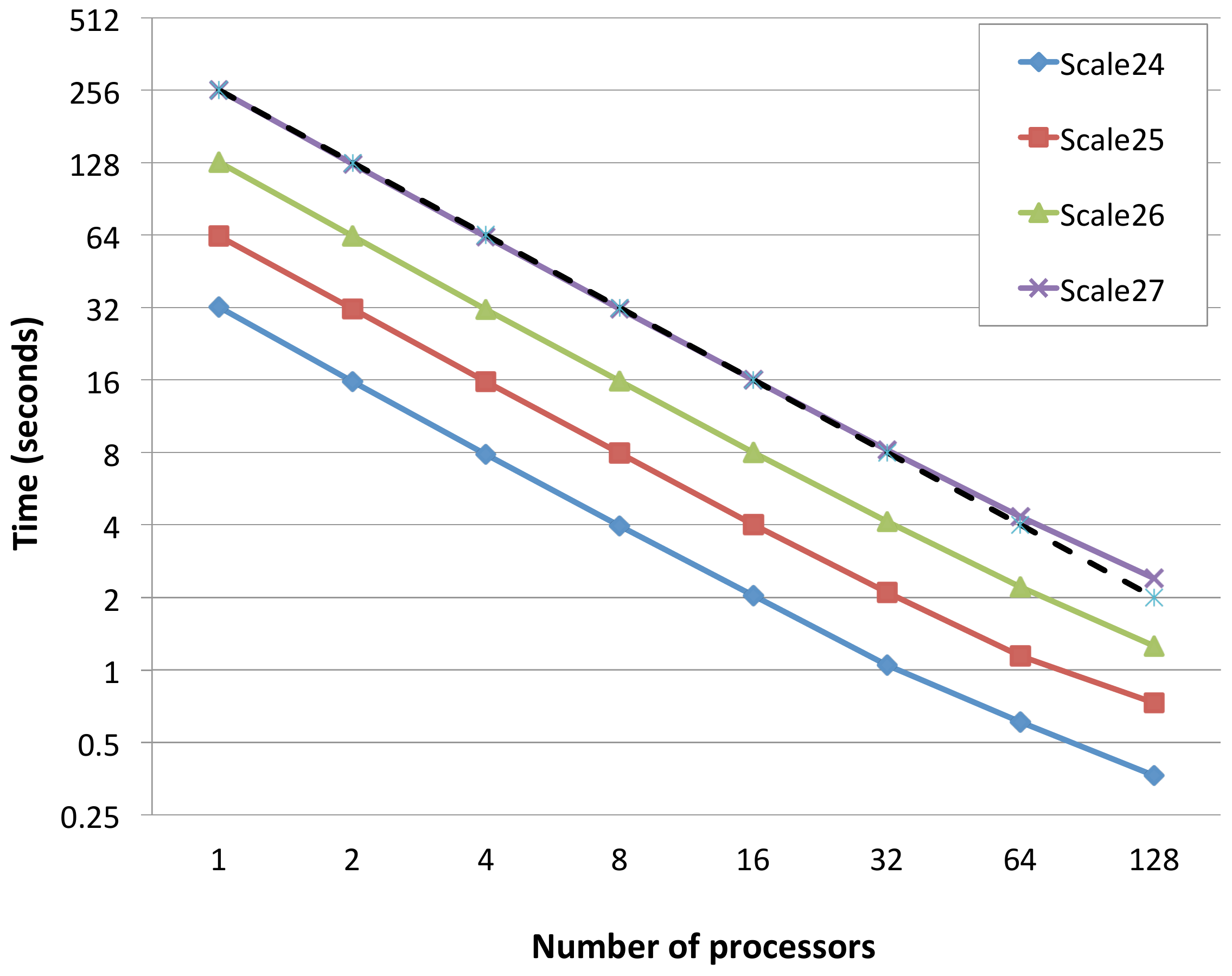} \label{f:xl-i-er}}
\subfigure[RMAT-ER (\textsc{DataflowRecursive})]{\includegraphics[width=0.47\textwidth]{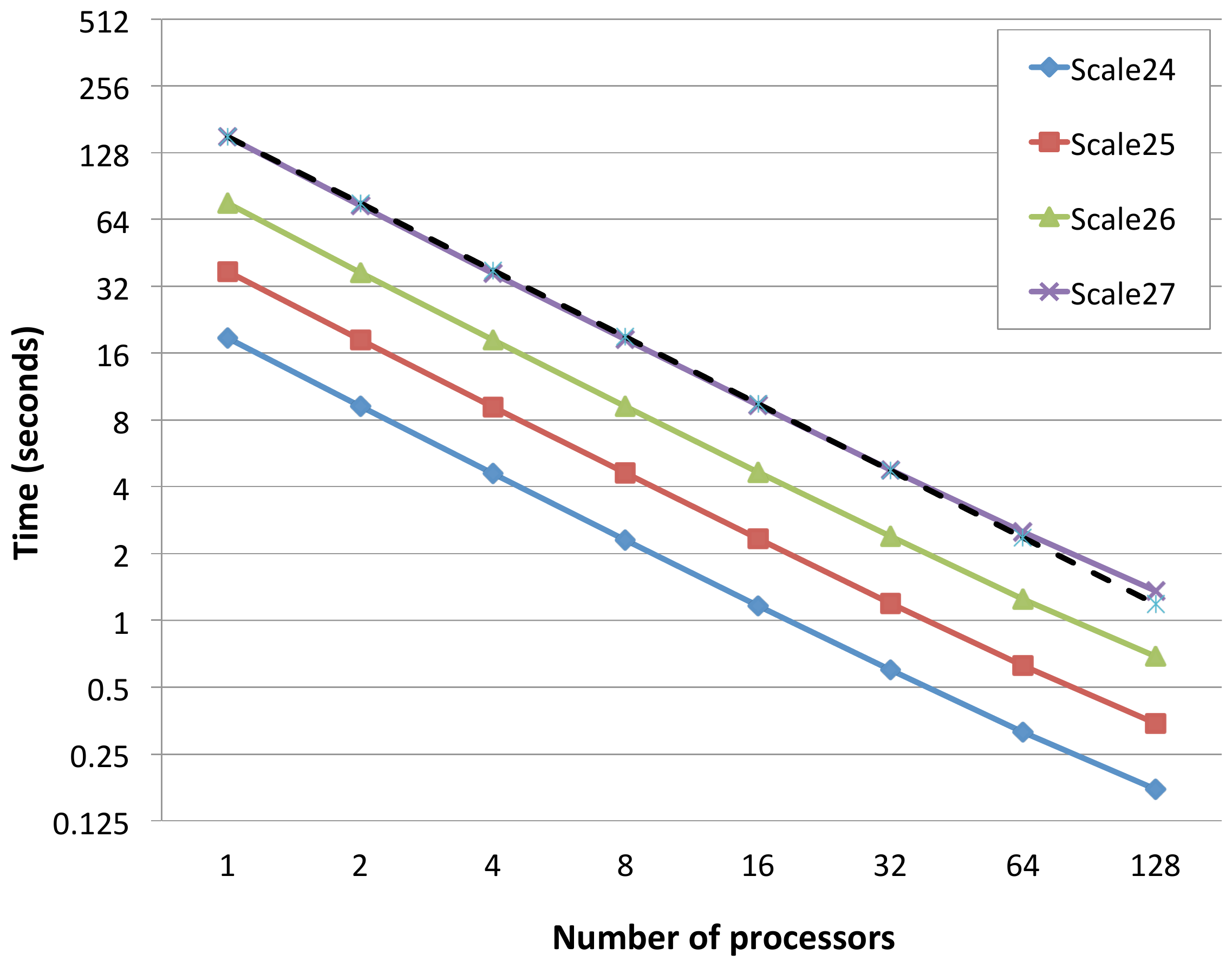} \label{f:xl-d-er}}
\subfigure[RMAT-G (\textsc{Iterative})] {\includegraphics[width=0.47\textwidth]{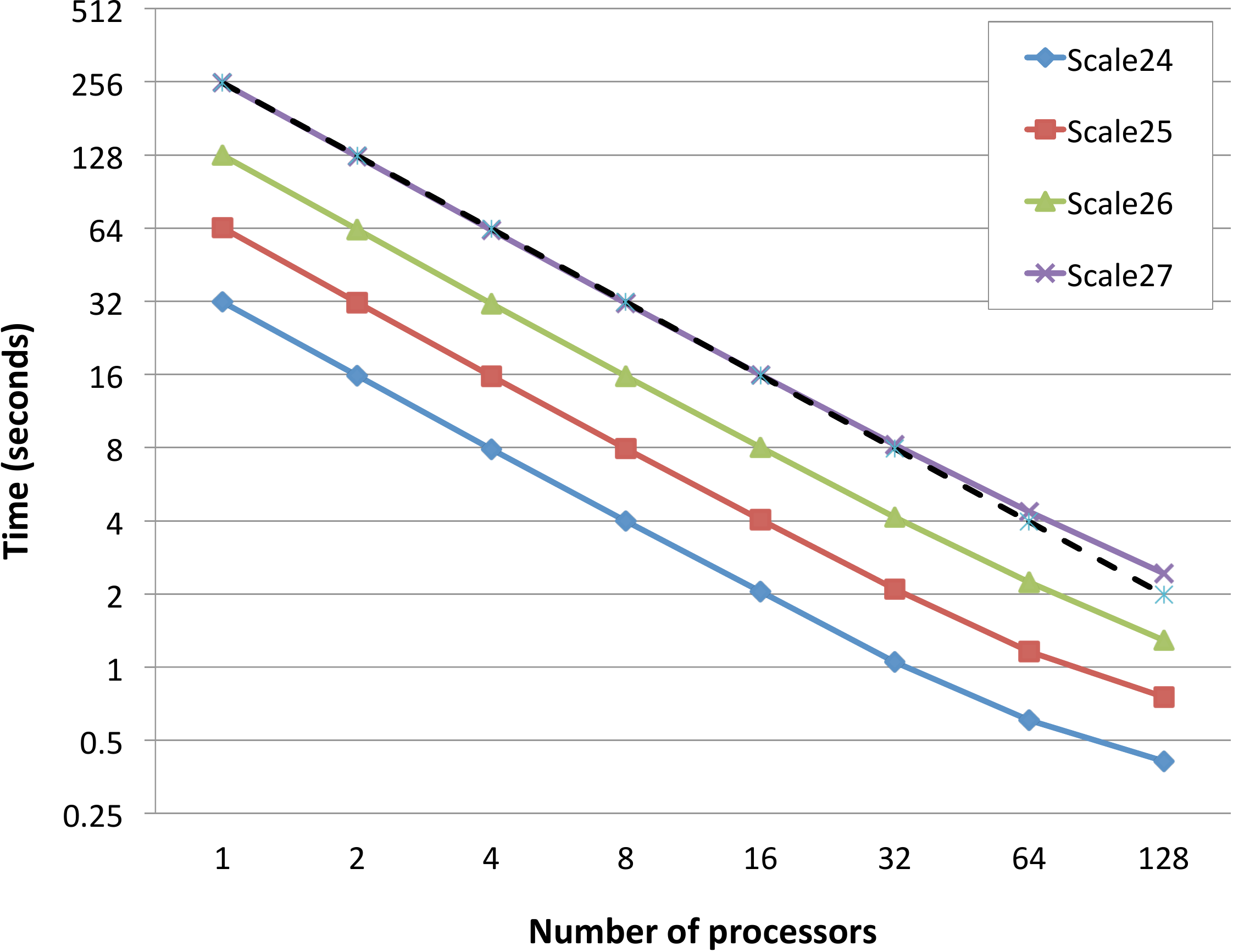} \label{f:xl-i-good}}
\subfigure[RMAT-G (\textsc{DataflowRecursive})] {\includegraphics[width=0.47\textwidth]{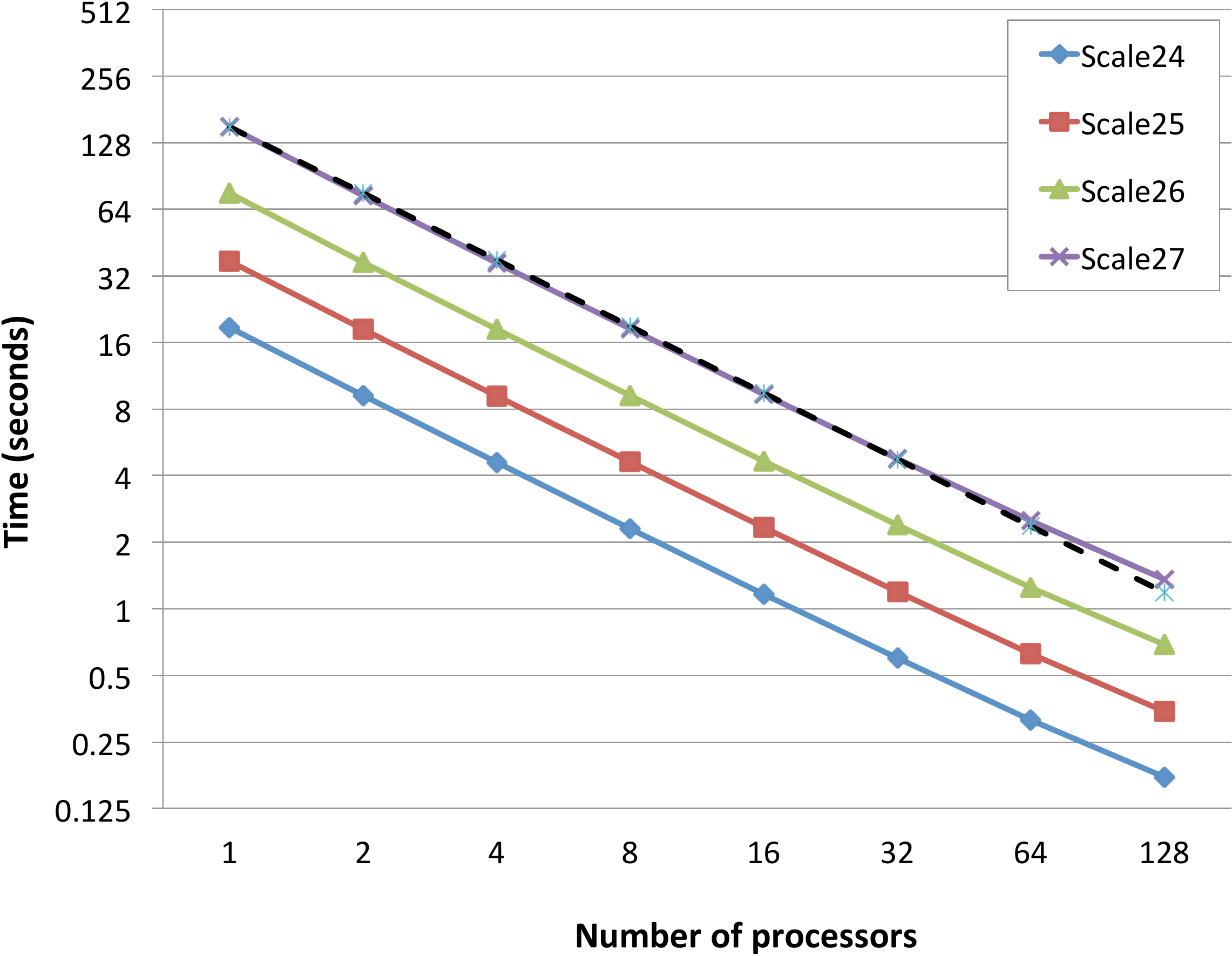} \label{f:xl-d-good}}
\subfigure[RMAT-B (\textsc{Iterative})] {\includegraphics[width=0.47\textwidth]{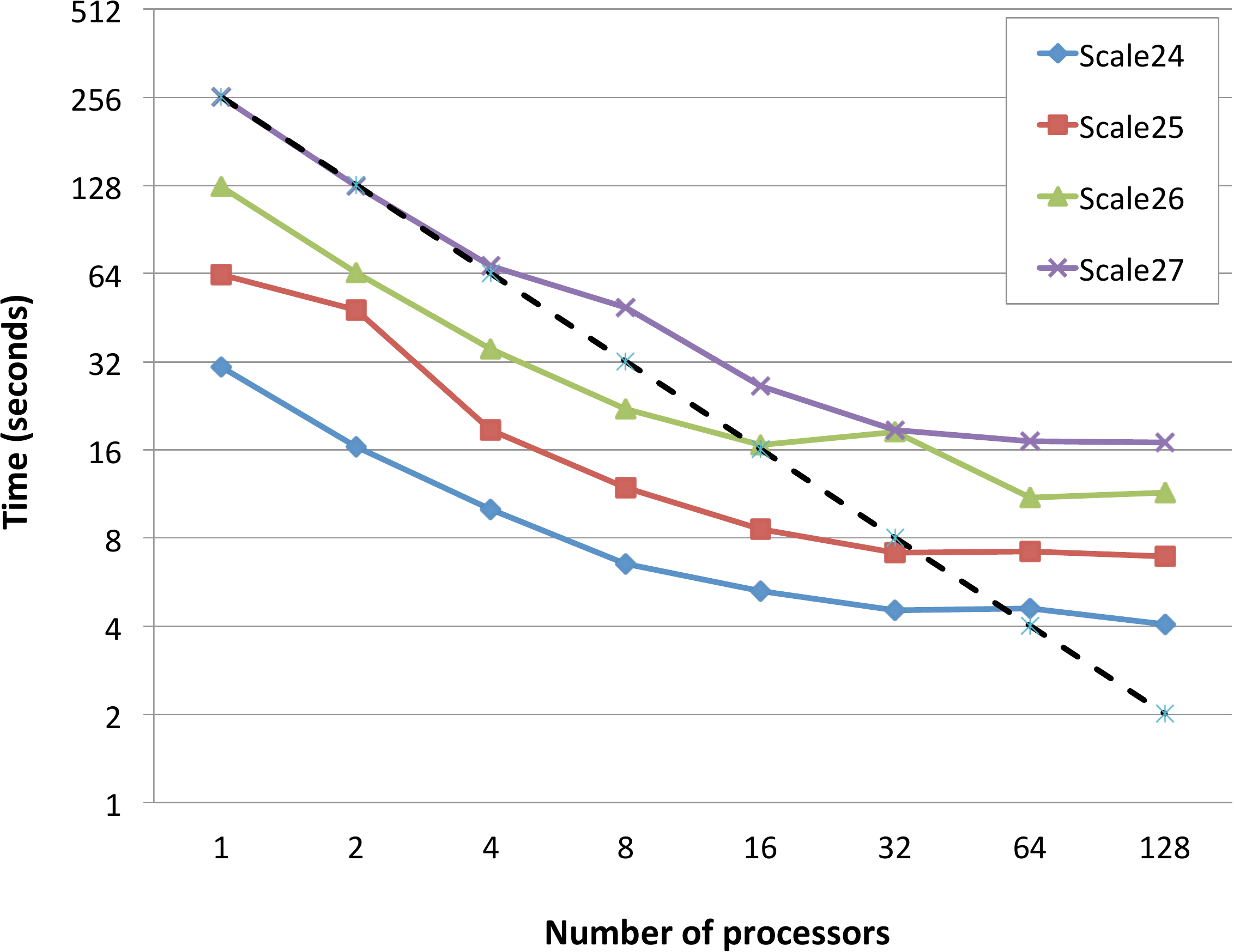} \label{f:xl-i-bad}}
\subfigure[RMAT-B (\textsc{DataflowRecursive})] {\includegraphics[width=0.47\textwidth]{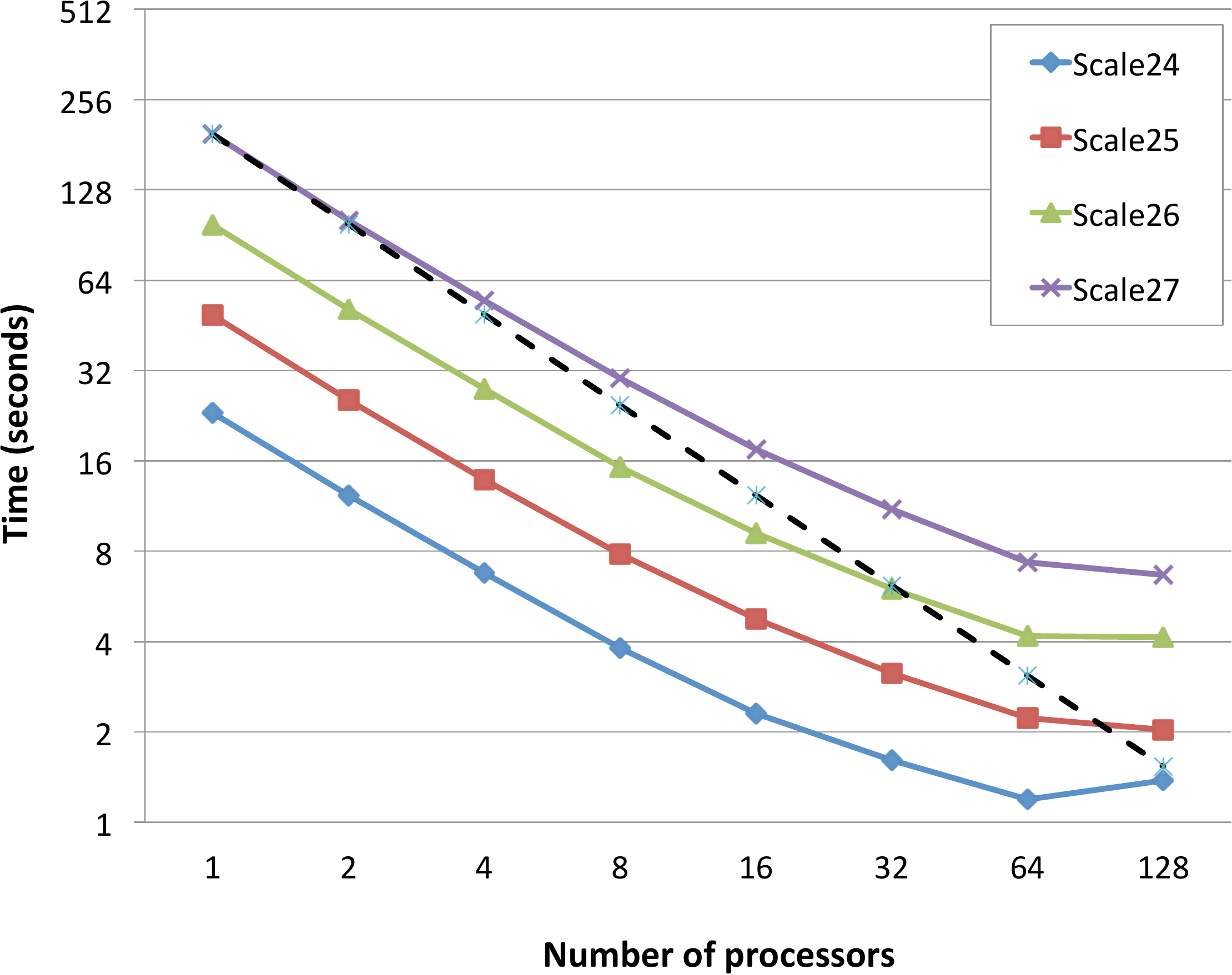} \label{f:xl-d-bad}}
\caption{\small Strong and weak scaling results on the Cray XMT for the graphs 
listed in Table~\ref{t:LargeScale}. 
For each run, a system request of maximum $100$ streams (threads) was placed. 
The dashed-line curve in each chart corresponds to {\em ideal} speedup 
on the largest graph (scale 27).
In all charts, both axes are in logarithmic scale.}
\label{f:xl}
\end{figure}

{\em Scalability results. }
The scalability results we obtained on the XMT with the maximum $100$ streams system request  
and using the large test graphs listed in Table~\ref{t:LargeScale} are summarized in Figure~\ref{f:xl};
the left set of figures shows results on algorithm \textsc{\textsc{Iterative}} and the right set of figures shows
results on algorithm \textsc{DataflowRecursive}.
In each case the ideal (linear) scaling of the {\em largest} (scale 27) graph is depicted by  the dashed line.

It can be seen that both \textsc{Iterative} and \textsc{DataflowRecursive} 
scale well to the maximum number of available processors (128) 
on the RMAT-ER and RMAT-G graphs; on the RMAT-B graph,
\textsc{DataflowRecursive} scales to 64 processors and  \textsc{Iterative} 
to 16 processors.
The poorer scalability of \textsc{Iterative} is likely due to the relatively large number of conflicts 
generated (and consequently, the relatively large number of iterations required) 
as a result of the massive concurrency utilized on the XMT in coloring the vertices.
It can also be seen that the runtime of \textsc{Iterative} is about twice that of \textsc{DataflowRecursive}.
This is mainly because of the conflict-detection phase in \textsc{Iterative}, 
which entails a second pass through the (sub)graph data structure, 
a phase that is absent in \textsc{DataflowRecursive}.

Note that Figure~\ref{f:xl} also shows {\em weak} scaling results. 
In each subfigure, observe that the input size is doubled as one goes 
from one input size (say 24) to the next (say 25) and the number of processors 
is doubled as one goes from one data point on the horizontal axis to the next.
Imagine running a horizontal line cutting through the four curves in each subfigure. 
In most of the figures such a line will intersect the four curves at points that correspond  
to a near-ideal weak scaling behavior.

\subsection{Scalability Comparison on the Three Architectures}
\label{sec:scalability-comp}

\begin{figure}
\centering
\subfigure[RMAT-ER]{\includegraphics[width=0.47\textwidth]{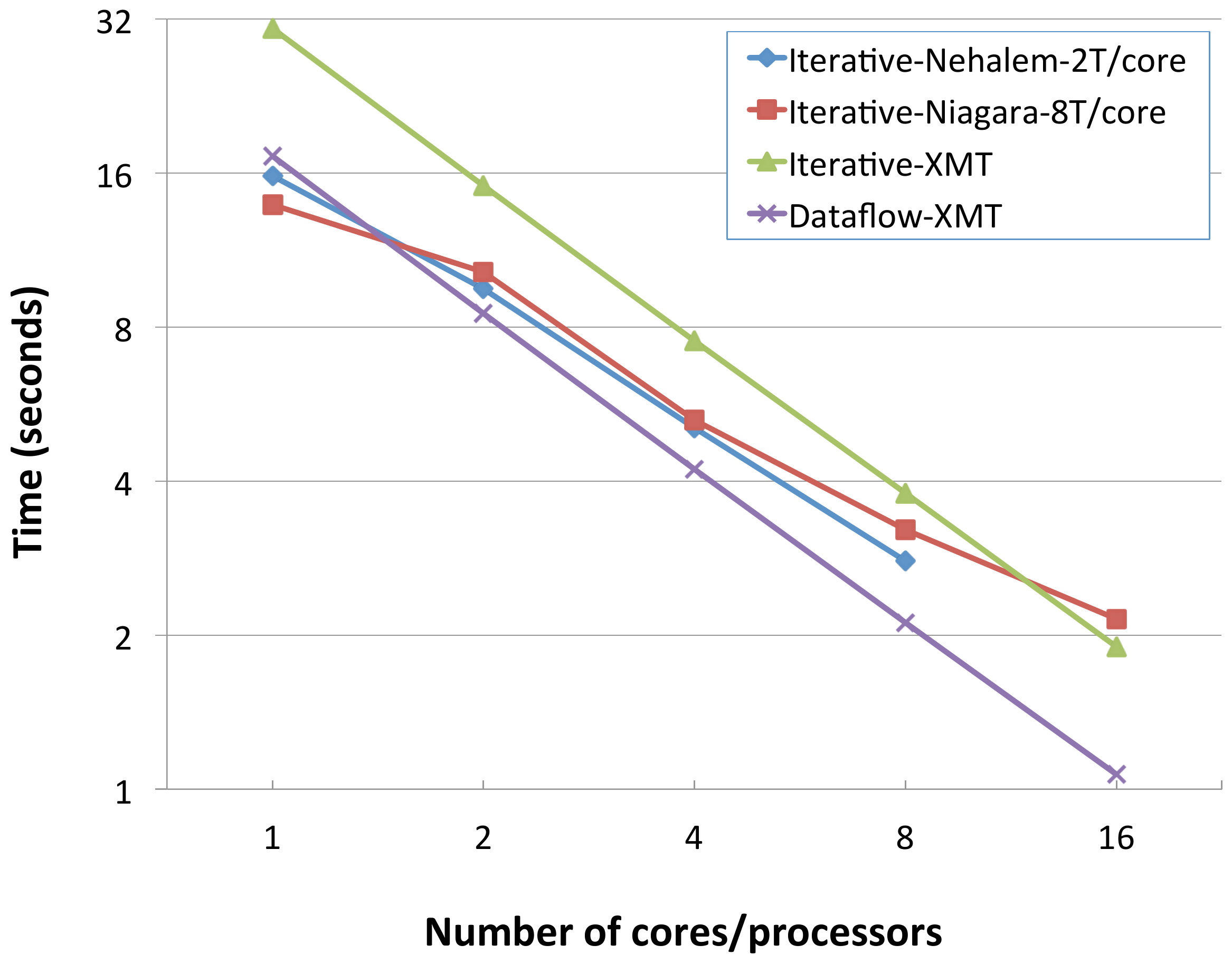} \label{f:all-er}}
\subfigure[RMAT-G] {\includegraphics[width=0.47\textwidth]{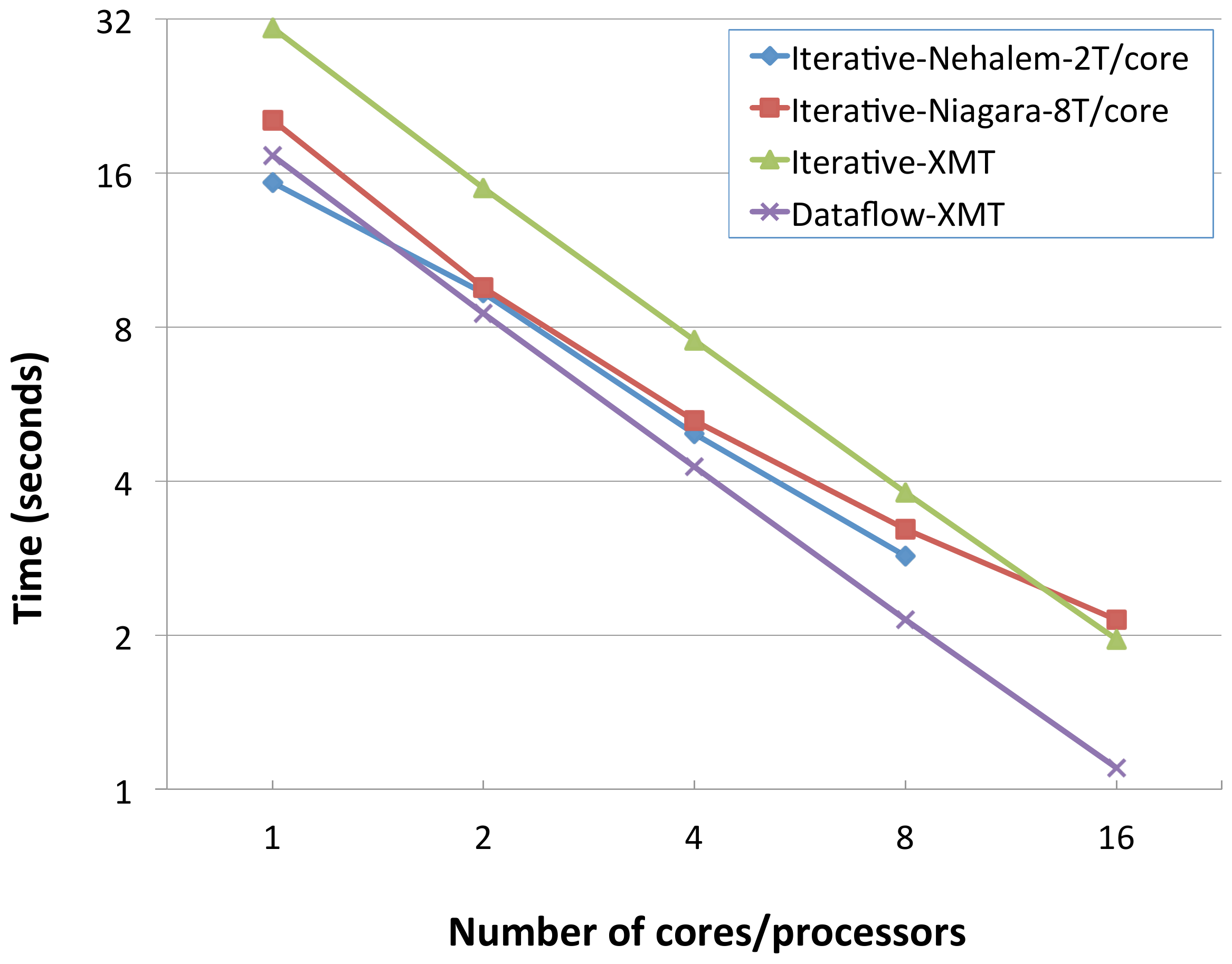} \label{f:all-good}}
\subfigure[RMAT-B] {\includegraphics[width=0.47\textwidth]{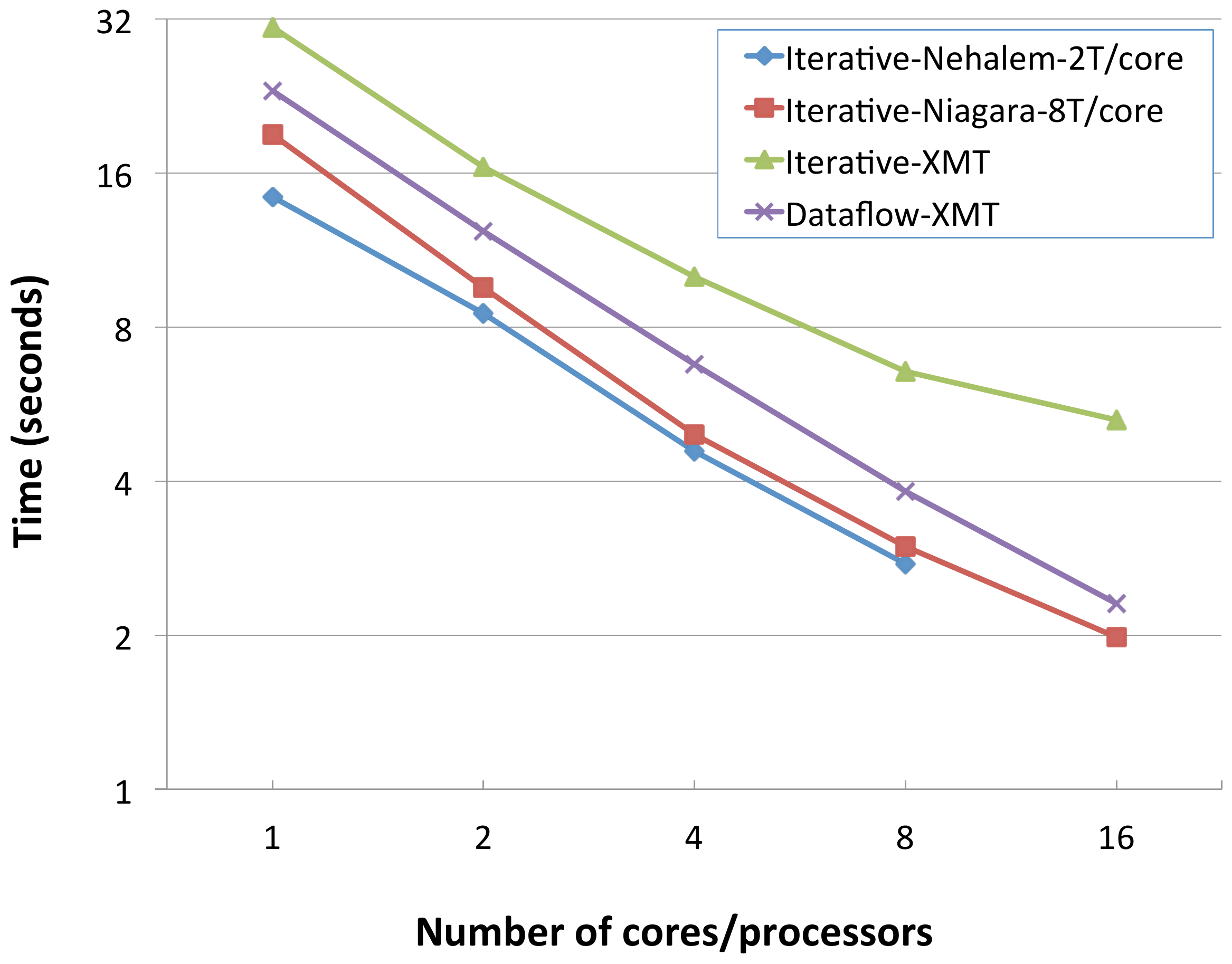} \label{f:all-bad}}
\caption{\small Strong scaling results on {\em all}  three platforms for the graphs listed in 
Table~\ref{t:prop}.
On the XMT, a system request was placed for maximum $100$ streams (threads).
In all charts, both axes are in logarithmic scale.}
\label{f:all}
\end{figure}

Figures~\ref{f:nehalem} and \ref{f:xl} have already shown some
of the differences in performance observed on the three architectures.
For further cross-platform comparison, we provide in
Figure~\ref{f:all} a condensed summary of strong scaling results on all three platforms at once. 
The three subfigures show results corresponding to runtimes in seconds on
the three test graphs of Table~\ref{t:prop}.
In each subfigure four runtime plots are shown, 
three corresponding to the performance of \textsc{Iterative} on the
three platforms and one corresponding to the performance of 
\textsc{DataflowRecursive} on the XMT.

To enable a common presentation on all three platforms,  we have shown 
in Figure~\ref{f:all} results on fewer processing units than what is available
on the Cray XMT. In particular, we show results when up to $8$ cores on the Nehalem, 
$16$ cores on the Niagara~2, and $16$ processors on the XMT are used.   
Note also that for a given number of cores (or processors in the case of XMT),
the amount of concurrency utilized and the type of 
multithreading (MT) employed varies from platform to platform. 
On the XMT, out of  the $128$ threads 
available per processor,  a system request for a maximum of $100$ was placed.
On the Niagara~2, the maximum eight threads available per core were used,
and on Nehalem the two available threads per core were used.   
This means the {\em total} amount of concurrency involved is a possible maximum of 
12,800 threads (with interleaved MT) on the XMT, 
exactly $128$ threads (with simultaneous MT) on the Niagara~2, and 
exactly $16$ threads  (with simultaneous MT) on the Nehalem.

On the XMT, it can again be seen that \textsc{DataflowRecursive} scales 
nearly ideally on RMAT-ER and RMAT-G, and quite decently on RMAT-B, the most difficult input. 
Algorithm \textsc{Iterative} scales in an analogous fashion on the same platform.
It can also be seen that \textsc{Iterative} scales well not only on the XMT, but
also on the Niagara~2 and Nehalem platforms, in that relative ordering.
The relative ordering is in part due  to the difference in the amount of thread concurrency
exploited, which is the highest on the XMT and the lowest on the Nehalem.

\subsection{\textsc{Iterative} and degree of concurrency}
\label{sec:iterative-more}

\begin{figure}
\centering
\subfigure[{\small Total number of conflicts}]{\includegraphics[width=0.45\textwidth]{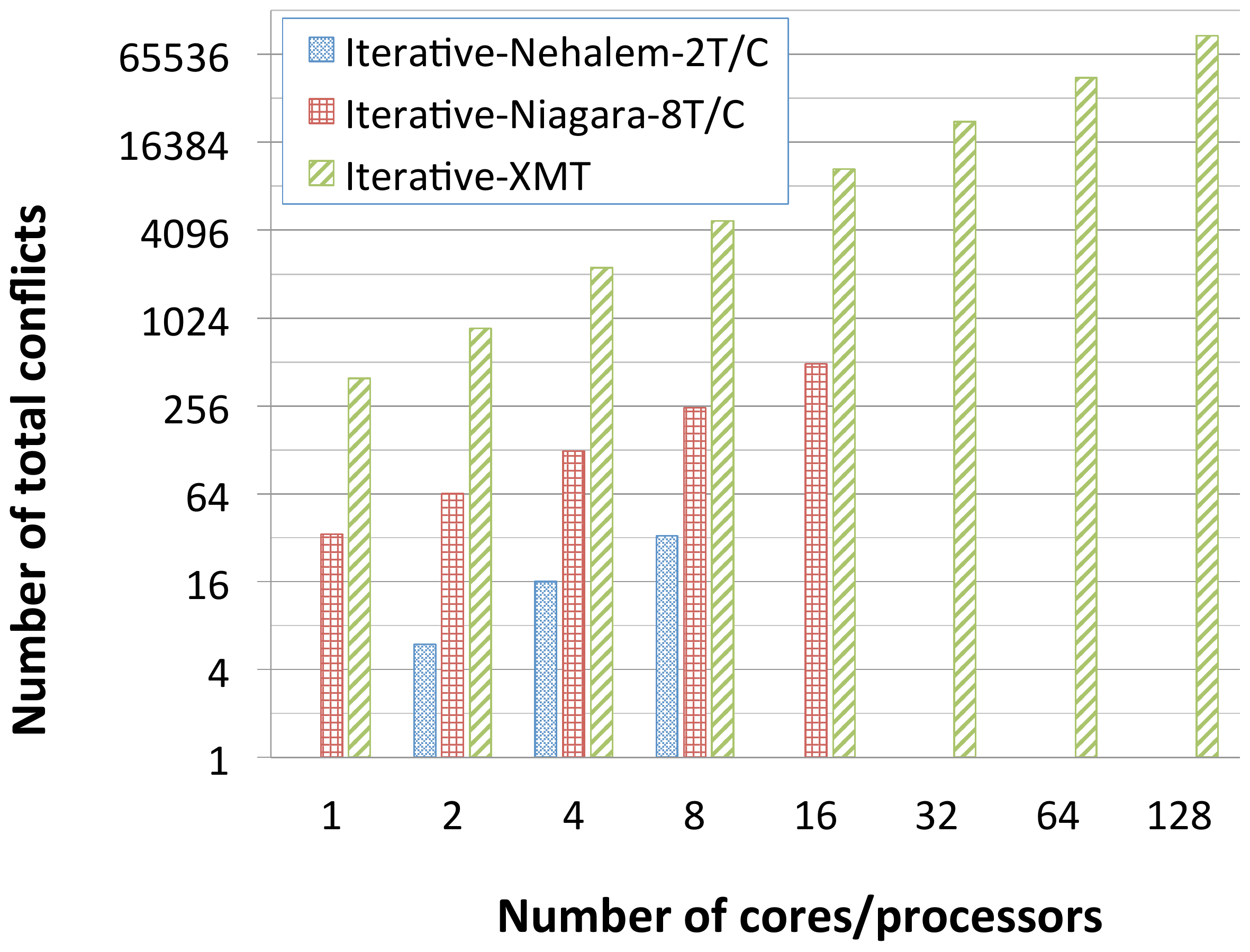} \label{f:conflicts}}
\subfigure[{\small Number of iterations}]{\includegraphics[width=0.45\textwidth]{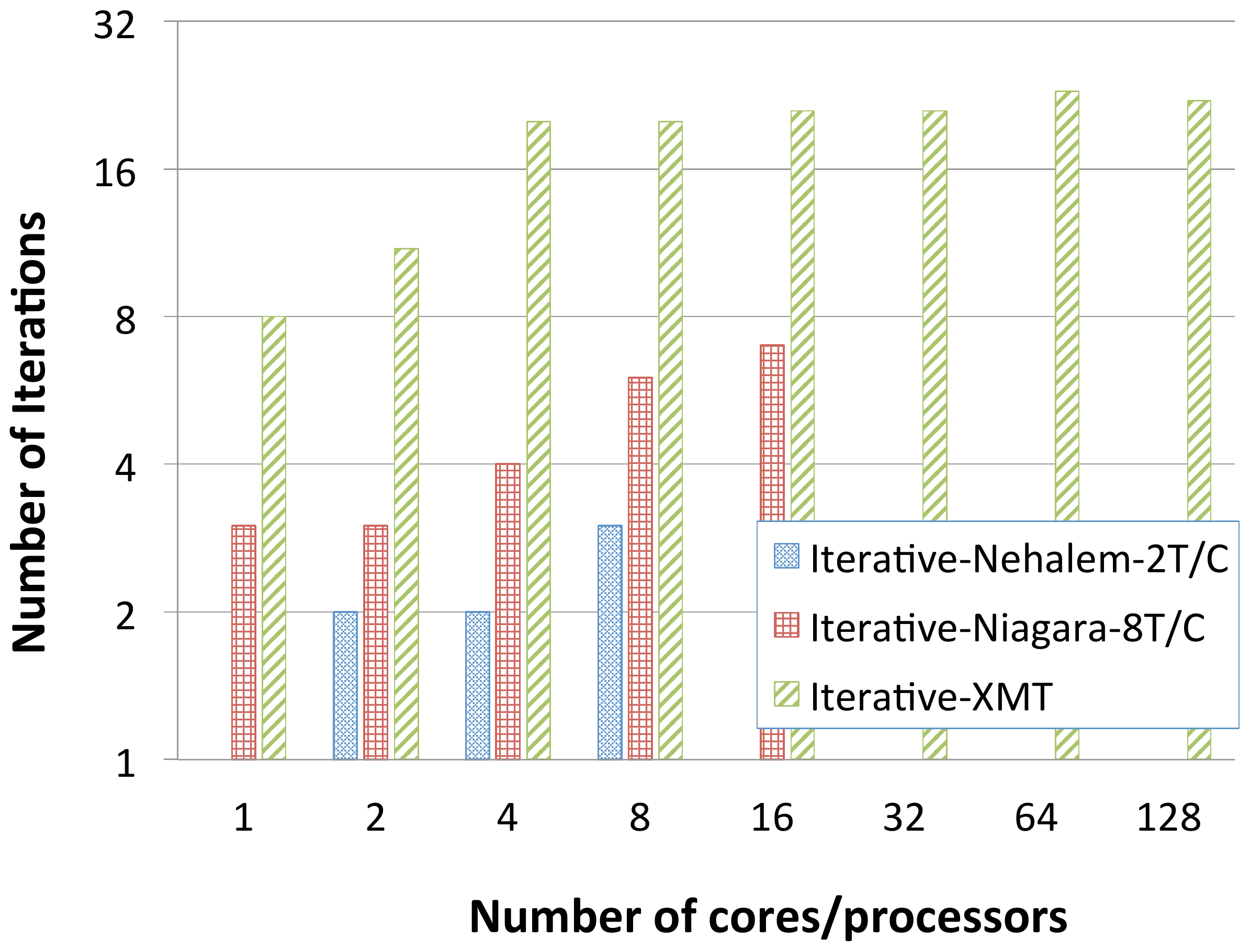} \label{f:iterations}}
\subfigure[{\small Number of conflicts per iteration}]{\includegraphics[width=0.45\textwidth]{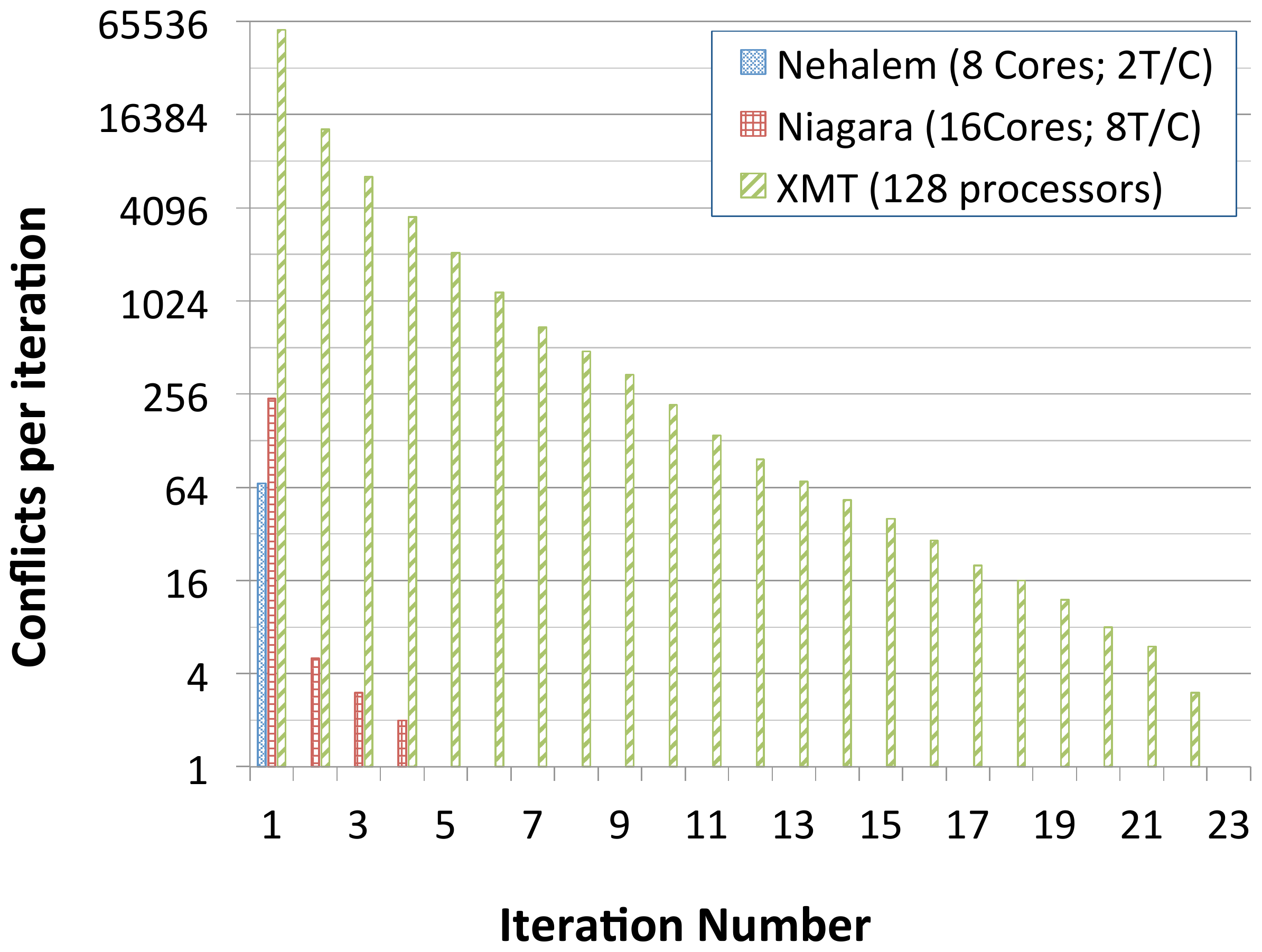} \label{f:conflictsvsiter}}
\caption{\small 
(a): Comparison of total number of conflicts generated in the \textsc{Iterative} algorithm
on the different platforms for the RMAT-B graph listed in Table~\ref{t:prop}.
(b): Total number of iterations the algorithm needed to complete on the same graph.
(c): Breakdown of number of conflicts per iteration.}
\label{f:conflicts-iterations}
\end{figure}

For \textsc{Iterative}, higher concurrency would also mean higher likelihood
for the occurrence of conflicts and, as a result, increase in the number of iterations
involved. Figure~\ref{f:conflicts} shows the {\em total} number of conflicts generated during the
course of algorithm \textsc{Iterative} on the three platforms for 
the most hostile of the three input types, the RMAT-B graph. 
As expected, the relative ordering in terms of number of conflicts
is Nehalem followed by Niagara~2 followed by the XMT. 
The number of these conflicts is seen to be small relative to the number of vertices
(16 million) in the graph. 
As Figure~\ref{f:iterations} shows, the total number of iterations the algorithm needed
to resolve these conflicts is quite modest, even on the XMT, 
where the number of conflicts is relatively large.
Figure~\ref{f:conflictsvsiter} shows how the total number of conflicts is divided across 
the iterations when $128$ processors on the XMT, $16$ cores of the Niagara 2 
(with $8$ threads per core) and $8$ cores of the Nehalem 
(with $2$ threads per core) are used. As this figure shows, 
about 90\% of the conflicts occur in the first iteration and 
the number of conflicts drops drastically in subsequent iterations.
This suggests that it might be worthwhile to switch to a sequential computation
once a small enough number of conflicts is reached, rather than proceed iteratively in parallel.
We will investigate this in future work.

\subsection{Number of colors}
\label{sec:colors}

The {\em serial} greedy coloring algorithm used $10$, $27$, and $143$ 
colors on the three graphs RMAT-ER, RMAT-G, and RMAT-B listed 
in Table~\ref{t:prop},  respectively. 
These numbers are much smaller than the 
maximum degrees (42; 1,278; and 38,143) in these graphs, 
which are upper bounds on the number of colors the greedy algorithm uses. 
Clearly, the gap between the upper bounds and the actual number of colors used is very large,
which suggests that the serial greedy algorithm is providing reasonably 
good quality solutions. Indeed experiments have shown that the greedy algorithm  
often uses near-optimal numbers of colors on many classes of graphs \cite{CoMo:83, colpack-acm}.  

\begin{figure}
\centering
\subfigure[RMAT-ER]{\includegraphics[width=0.45\textwidth]{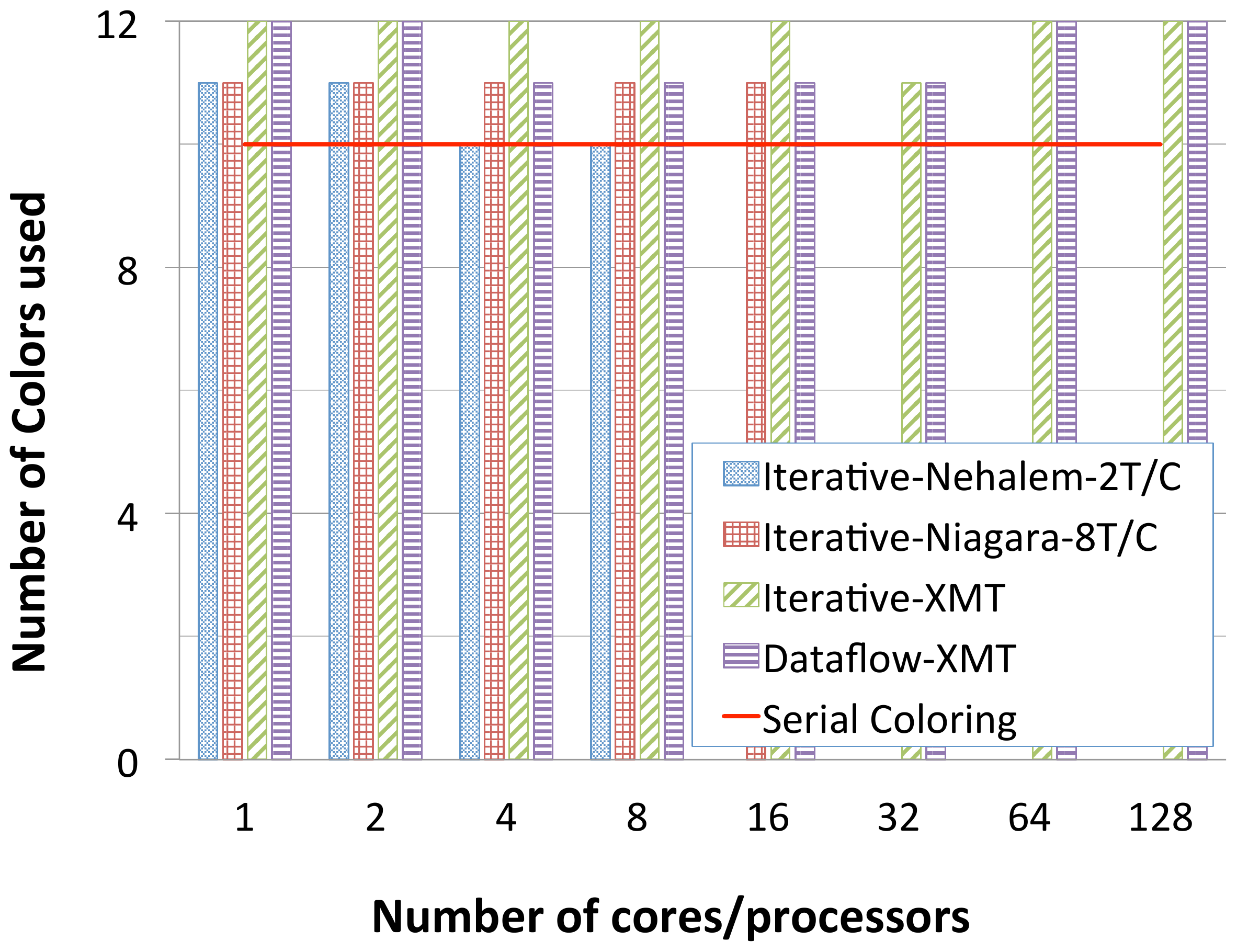} \label{f:all-er-c}}
\subfigure[RMAT-G]{\includegraphics[width=0.45\textwidth]{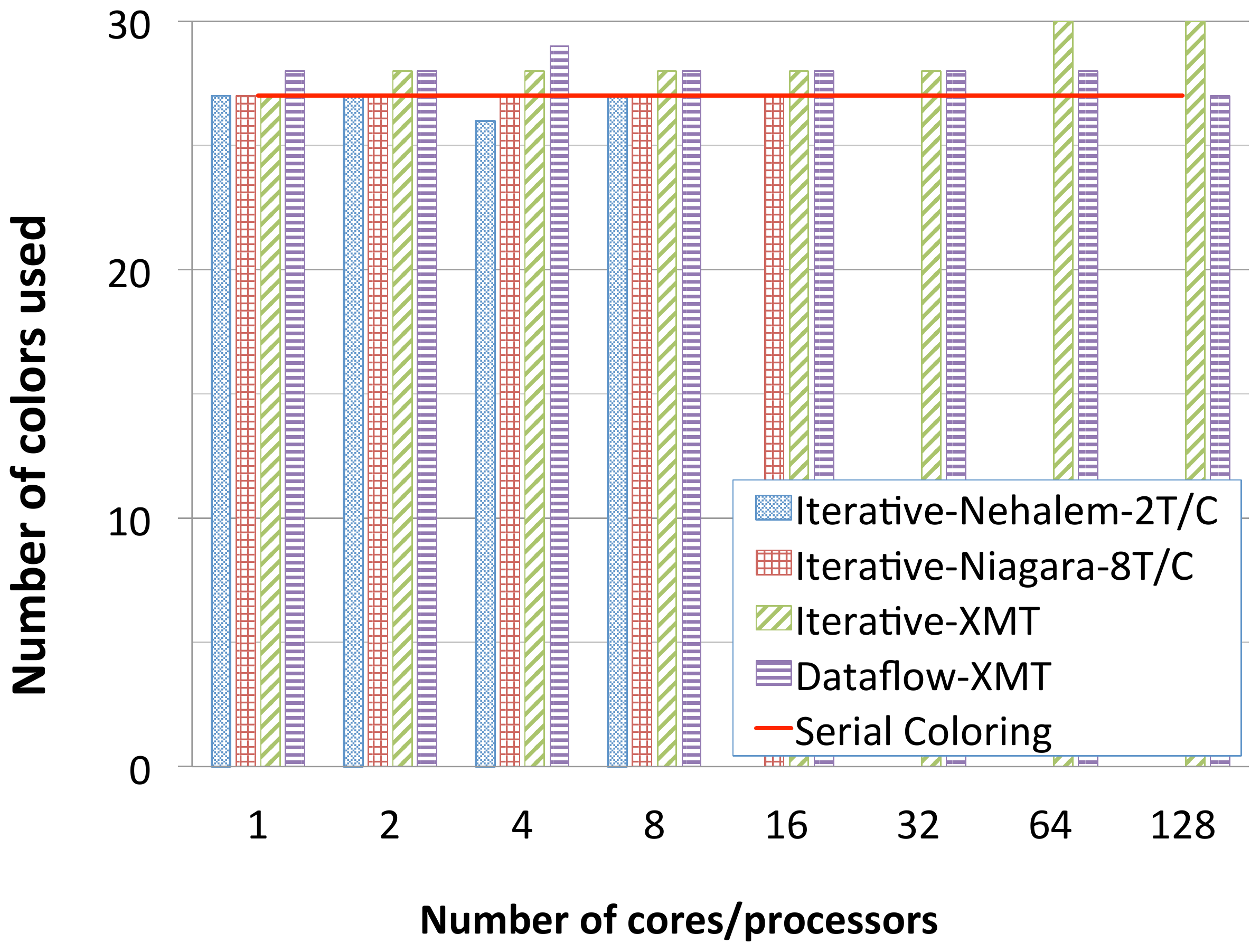} \label{f:all-good-c}}
\subfigure[RMAT-B]{\includegraphics[width=0.45\textwidth]{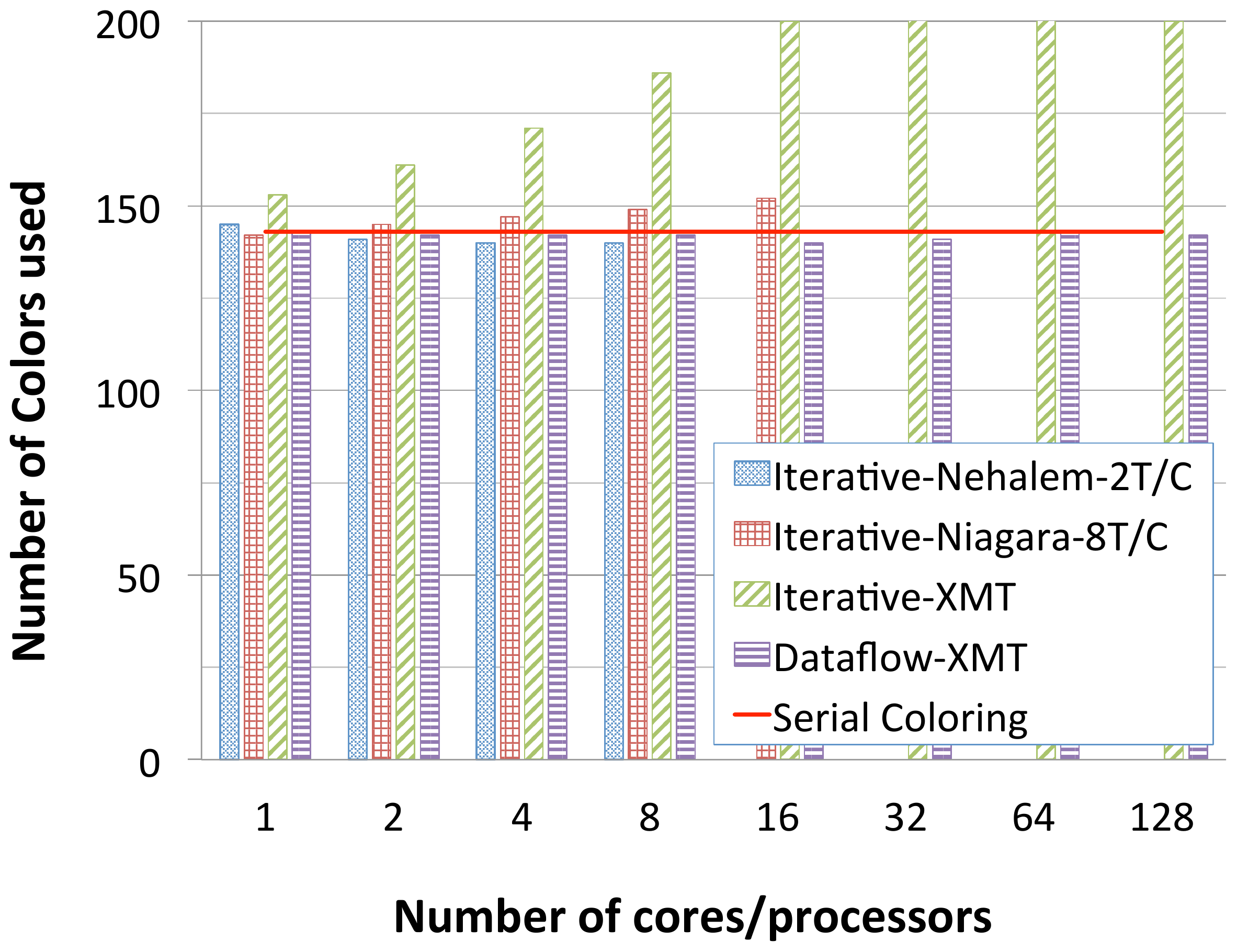} \label{f:all-bad-c}}
\caption{\small Number of colors used by the two multithreaded algorithms on the various 
architectures for the graphs listed in Table~\ref{t:prop}. 
The number used by the {\em serial} greedy algorithm is depicted with the horizontal red line.}
\label{f:all-colors}
\end{figure}

Figure~\ref{f:all-colors} summarizes  the number of colors the multithreaded algorithms 
\textsc{Iterative} and \textsc{DataflowRecursive} use on the three test graphs 
as the number of cores/processors on the various platforms is varied. 
It can be seen that both algorithms use about the same number of colors as the serial greedy algorithm
and the increase in number of colors with an increase in concurrency is 
none to modest---only for \textsc{Iterative} on the XMT 
(where concurrency is massive) and the graph RMAT-B do we see a modest increase
in number of colors as the number of processors is increased.

\section{Conclusion}
\label{sec:conc}

We presented a heuristic multithreaded algorithm for graph coloring 
that is suitable for any shared-memory system, including multi-core platforms.
The key ideas in the design of the algorithm are {\em speculation} and {\em iteration}.  
We also presented a massively multithreaded algorithm for coloring designed using
{\em dataflow} principles exploiting the fine-grain, hardware-supported 
synchronization mechanisms available on the Cray XMT. 
Using a carefully chosen set of input graphs, covering a wide range of problem types, 
we evaluated the performance of the algorithms on three different platforms---Intel Nehalem, 
Sun Niagara~2, and Cray XMT---that feature varying degrees of 
multithreading and caching capabilities to tolerate latency.
The iterative algorithm (across all three architectures) and 
the dataflow algorithm (on the Cray XMT)
achieved near-linear speedup on two of the three input graph classes considered and 
moderate speedup on the third, most difficult, input type.  
These runtime performances were achieved without compromising the quality of the solution
(i.e., the number of colors) produced by the underlying serial algorithm.
We also characterized the input graphs and provided insight on bottlenecks for performance. 

Backed by experimental results, some of the more general conclusions we draw about
parallel graph algorithms include:
\begin{itemize}
\item simultaneous multithreading provides an effective way to tolerate latency, as can be seen from  
the experiments on the Nehalem and Niagara~2 platforms where the runtime 
using $N$ threads on one core is found to be similar to the runtime using one thread on $N$ cores, 
\item the impact of lower clock frequency and smaller cache memories can be ameliorated with a greater thread concurrency, 
as can be seen in the better performance obtained on the XMT and the Niagara~2 relative to the Nehalem,
\item when supported by light-weight synchronization mechanisms in hardware, 
parallelism should be exploited at fine grain, and
\item graph {\em structure} critically influences the performance of parallel graph algorithms.
\end{itemize}
We expect these insights to be helpful in the design of high performance algorithms for
irregular problems on the impending many-core architectures. 

\section*{Acknowledgments}
We thank the anonymous referees and Fredrik Manne for their valuable 
comments on an earlier version of the manuscript. This research was supported by the 
U.S. Department of Energy through the CSCAPES Institute (grants
DE-FC02-08ER25864 and DE-FC02-06ER2775), by the National Science
Foundation through grants CCF-0830645, CNS-0643969, OCI-0904809,
and OCI-0904802, and by the Center for Adaptive Supercomputing
Software (CASS) at the Pacific Northwest National Laboratory. The
Pacific Northwest National Laboratory is operated by Battelle for the
U. S. Department of Energy under Contract DE-AC06-76L01830.

\bibliographystyle{plain}
\bibliography{MTcoloring}

\end{document}